 \documentclass[journal]{IEEEtran}
\usepackage{cite}
\usepackage{amsmath,amssymb,amsfonts}
\usepackage{algorithm}  
\usepackage{algpseudocode}  
\usepackage{graphicx}
\usepackage{epstopdf}
\usepackage{textcomp}
\usepackage{xcolor}
\usepackage{bm}
\usepackage{enumerate}
\usepackage{subcaption}
\usepackage{caption} 
\usepackage{amsmath}

\ifCLASSINFOpdf
\else
\fi
\begin{document}
	
\title{Mobile Tracking via Target-Mounted 

IRS-Assisted ISAC System}
% note locations of commas and nonbreaking spaces ( ~ ) LaTeX will not break
%\author{authors}
% \author{Ziheng~Zhang,~Wen~Chen,~\IEEEmembership{Senior~Member,~IEEE},~Zhendong~Li,~Xusheng~Zhu,~Qingqing~Wu,~\IEEEmembership{Senior~Member,~IEEE},\\~and~Jinhong~Yuan,~\IEEEmembership{Fellow,~IEEE\vspace{-2em}}

% 
\author{Ziheng~Zhang,~Wen~Chen,~\IEEEmembership{Senior~Member,~IEEE},~Qingqing~Wu,~\IEEEmembership{Senior~Member,~IEEE},~Zhendong~Li,~Qiong~Wu,~\IEEEmembership{Senior~Member,~IEEE},~Ming-Min~Zhao,~Wei~Feng~\IEEEmembership{Senior~Member,~IEEE
}
\vspace{-36pt}

% \thanks{This work is supported by National key project 2020YFB1807700, NSFC 62071296, Shanghai Kewei 22JC1404000. This work is supported by NSFC 62371289, NSFC 62331022 and ZTE Industry-University-Institute Cooperation Funds under Grant No. IA20240420003. This work is supported in part by NSFC 62401448 and National Postdoctoral Program for Innovative Talent BX20240277.}
% \thanks{This work is supported by National key project 2020YFB1807700, NSFC 62071296, Shanghai Kewei 22JC1404000. Wu's work is supported by NSFC 62371289 and NSFC 62331022.}

\thanks{Z. Zhang, W. Chen, and Qingqing Wu are with the Department of Electronic Engineering, Shanghai Jiao Tong University, Shanghai 200240, China (e-mail: zhangziheng@sjtu.edu.cn; wenchen@sjtu.edu.cn; 
qingqingwu@sjtu.edu.cn;)}
\thanks{Z. Li is with the School of Information and Communication Engineering, Xi'an Jiaotong University, Xi'an 710049, China (e-mail: lizhendong@xjtu.edu.cn).}
\thanks{Qiong Wu is with the School of Internet of Things Engineering, Jiangnan University, Wuxi 214122, China (e-mail: qiongwu@jiangnan.edu.cn).}
\thanks{Ming-Min Zhao is with the College of Information
Science and Electronic Engineering, Zhejiang University, Hangzhou 310027, China (e-mail: zmmblack@zju.edu.cn;)}
\thanks{Wei Feng is with the Department of Electronic Engineering, State Key Laboratory of Space Network and Communications, Tsinghua University, Beijing 100084, China (e-mail: fengwei@tsinghua.edu.cn)}

% \thanks{Nan Cheng is with the School of Telecommunications Engineering, Xidian University, Xi’an
% 710071, China (e-mail: dr.nan.cheng@ieee.org).}
% <-this % stops a space
%\thanks{K. Wang is with the School of Information Science and Technology, ShanghaiTech University, Shanghai 201210, China, and also with the School of Communication and Electronic Engineering, East China Normal University, Shanghai 200241, China (e-mail: wangkl2@shanghaitech.edu.cn).}

%\thanks{This work is supported by ...}
\thanks{(\emph{Corresponding author: Wen Chen.})}}
\maketitle
\begin{abstract}
This paper proposes a target-mounted intelligent reflecting surface (IRS)-assisted integrated sensing and communication framework for real-time unmanned aerial vehicle (UAV) tracking, addressing challenges such as link blockage and weak radar cross section in the low-altitude economy. By integrating the IRS onto the UAV, the system creates a mobile cooperative target that provides controllable line-of-sight echoes for self-tracking while acting as a mobile relay for ground communication enhancement. We establish a comprehensive three dimensions state evolution model for the maneuvering UAV. Based on this model, an extended Kalman filter is immediately implemented to achieve real time tracking of the moving UAV. To characterize the fundamental theoretical limits of this recursive estimation process, we derive the analytical posterior Cramer Rao bound and a closed form expression for the elliptical tradeoff performance bound to quantify the relationship between sensing precision and communication throughput. To ensure millisecond level responsiveness, we develop a low complexity joint beamforming design. By utilizing the analytical mapping between tracking and communication requirements, the proposed scheme yields closed form solutions for beamforming vectors, effectively bypassing the time consuming numerical iterations of conventional methods. Numerical simulations demonstrate that the proposed framework significantly outperforms traditional fixed-deployment benchmarks across complex maneuvering trajectories, achieving centimeter-level accuracy while substantially reducing transmit power and processing latency.
\end{abstract}


% \fi
\IEEEpeerreviewmaketitle

\section{Introduction}
Recently, the low-altitude economy (LAE) has emerged as a key economic driver. This sector is characterized by the rapid growth of unmanned aerial vehicles (UAVs) in various applications, such as logistics delivery, emergency rescue, and infrastructure inspection \cite{10955337,10906066,11359737}. To ensure flight safety and operational efficiency, integrated sensing and communication (ISAC) is essential. This technology enables both high-rate data transmission and precise target tracking within a unified hardware platform \cite{11072035,10879807}. However, implementing ISAC in low-altitude scenarios faces several critical challenges. In urban or mountainous areas, the direct line-of-sight (LoS) links are often blocked by buildings or trees, creating ``blind zones" for both sensing and communication \cite{11205853}. Moreover, most low-altitude targets have a very small radar cross section (RCS). This results in extremely weak echo signals that are difficult for the base station to detect \cite{10950390}. These issues lead to unstable connectivity and poor tracking performance \cite{10977743}. Therefore, overcoming these obstacles to enhance the overall performance of ISAC systems has become a top priority.

To address the aforementioned challenges, intelligent reflecting surface (IRS) has emerged as a promising technology to reconfigure the wireless propagation environment \cite{10443608}. By synergistically adjusting the phase shifts of a large number of passive reflecting elements, IRS can establish virtual LoS links to bypass physical obstacles and significantly enhance signal quality \cite{10160136,9650755,9606864}. In ISAC systems, the use of IRS to improve spectral efficiency and expand coverage has become a dominant trend \cite{10497119,10440056}. Extensive research has focused on IRS-assisted joint beamforming designs aimed at coordinating resource competition between sensing and communication tasks \cite{9769997,9591331}. Furthermore, IRS has demonstrated great potential in enhancing sensing accuracy and target resolution by providing additional spatial degrees of freedom \cite{10527368}. Specific explorations into secure transmission \cite{10143420}, energy efficiency optimization \cite{10005150}, and precise multi-IRS target localization \cite{10475369,10218356} within complex scenarios have also been reported. These research findings establish a solid theoretical foundation for the wide application of IRS-ISAC systems. However, as the focus shifts toward high-mobility tracking in the low-altitude economy, existing technical solutions still face severe tests in terms of deployment architecture and theoretical depth.

Despite the significant progress in IRS-assisted ISAC systems, a major limitation of current research lies in the pervasive assumption of fixed IRS deployment on static infrastructures such as building surfaces or roadside towers \cite{10141975}. While these stationary configurations are effective for enhancing regional coverage, they struggle to support the high-mobility requirements of the emerging low-altitude economy \cite{10632049}. Specifically, the limited reflection beamwidth of a fixed IRS makes it difficult to provide continuous and stable LoS links as as a UAV moves through wide and complex areas. This unavoidably results in frequent beam switches and increased signaling overhead \cite{10056405}. Furthermore, existing ISAC frameworks are mainly based on ‌two-dimensional (2D) geometric models or simplified static parameter estimation and  these frameworks fail to capture the complex dynamic nature of mobile targets \cite{10490002}. There is a notable lack of comprehensive 3D state evolution modeling that integrates not only position but also real-time velocity and acceleration, which is vital for achieving ultra-reliable tracking performance in safety-critical low-altitude applications \cite{10506632}. Without a robust 3D motion aware framework, the ability of IRS to reduce fast channel fluctuations and handle the large geometric changes in UAV tracking cases remains poorly used.

Another critical research gap in current literature lies in the lack of fundamental analytical insights and the high computational complexity of existing joint beamforming designs. Current research efforts primarily formulate IRS-assisted ISAC resource allocation as complex non-convex optimization problems. These problems are typically solved using numerical iterative techniques such as semi-definite relaxation (SDR), successive convex approximation (SCA), and majorization-minimization (MM) \cite{10052711}. While these numerical methods are effective for finding local optima in quasi-static scenarios, they fail to provide clear closed-form expressions that characterize the fundamental Pareto frontier between sensing precision and communication reliability \cite{9844707}. The absence of such theoretical basis makes it difficult for the system to adaptively adjust its beamforming strategy in real-time response to the instantaneous geometric variations between the base station and the mobile target \cite{9416177}. Furthermore, the tight coupling between active transmit beamforming at the BS and passive phase-shift design at the IRS leads to a massive search space, requiring a significant number of iterations to achieve convergence \cite{9416177}. In high mobility applications, the millisecond scale evolution of a UAV state demands immediate responsiveness for stable tracking. Conventional iterative algorithms, however, impose heavy computational loads and processing delays that are incompatible with strict real time requirements. Specifically, rapid target motion during computation intervals delays beam updates, resulting in severe pointing mismatches \cite{10577673}. This latency between the optimized beam and the actual target position results in a sharp drop in tracking accuracy or even a total loss of connection \cite{10411853}. Consequently, developing a theoretical framework that yields closed-form performance bounds and low-complexity optimization algorithms is essential for enabling efficient and reliable operations in mobile IRS-assisted ISAC systems.

Motivated by the coverage limitations of stationary IRS and the high computational latency of existing iterative algorithms, this paper proposes a target-mounted IRS-assisted ISAC framework for real-time UAV tracking. By integrating the IRS onto the UAV, the system creates a mobile cooperative target that provides controllable LoS echoes for sensing and acts as a mobile relay for communication. We establish a 3D state evolution model and derive the posterior Cramér-Rao bound (PCRB) for tracking analysis. To ensure millisecond-level responsiveness, a low-complexity beamforming design based on case analysis is developed to provide closed-form solutions that satisfy tracking and quality-of-service requirements with minimal power. Finally, the Pareto frontier is analyzed to quantify the optimal trade-off between sensing and communication functions under various 3D geometric configurations. The main contributions of the paper are summarized as follows.
\begin{itemize}
\item  \textbf{First, we establish a comprehensive three-dimensional state evolution model tailored for high-mobility unmanned aerial vehicle tracking assisted by a target-mounted intelligent reflecting surface.} In contrast to existing two-dimensional or static integrated sensing and communication frameworks, our model explicitly captures the real-time coupling of the target position and velocity vectors in three-dimensional space. Furthermore, an extended Kalman filter (EKF) is implemented to achieve real-time tracking of the moving UAV.

{\color{blue}\item \textbf{Second, we analytically characterize the fundamental limits of the ISAC system by deriving its exact Pareto performance frontier.} Through spatial subspace projections, we establish a rigorous closed-form elliptical boundary equation. This explicitly quantifies the fundamental trade-off between 3D tracking precision and communication throughput, defining the absolute theoretical performance ceiling of the target-mounted architecture without relying on heuristic numerical approximations.
    
\item \textbf{Finally, guided by the derived analytical boundary, we develop an ultra-low-complexity closed-form joint beamforming algorithm to strictly guarantee real-time responsiveness.} By operating directly on the exact Pareto frontier via case analysis, the proposed design entirely bypasses the massive iterations and relaxation gaps inherent in conventional numerical solvers. Extensive results confirm that it delivers a massive computational acceleration while flawlessly achieving the theoretical global optimum for high-mobility tracking.}

\end{itemize}

The rest of this paper is organized as follows. In Section II, we introduce the system model for target-mounted IRS-assisted integrated sensing and communication and formulate the transmit power minimization problem. In Section III, we present the three-dimensional state evolution and the theoretical derivation of the PCRB. Section IV presents the proposed joint beamforming design, where we first derive the analytical elliptical performance boundary to characterize the optimal trade-off. Based on this, a case-analysis algorithm is developed to achieve Pareto optimality with low computational complexity. In Section V, extensive simulation results are provided to validate the superiority of the proposed framework and to analyze the Pareto frontier under various geometric configurations. Finally, Section VI concludes the paper.

\textit{Notations:} Matrices, vectors and scalars
are represented by bold uppercase, bold lowercase and standard lowercase letters, respectively. For a complex-valued scalar $x$, $\left| {x} \right|$ denotes its absolute value. For a complex-valued vector $\bf{x}$, ${{\left\| \bf{x}  \right\|}_{p}}$, ${{\left[ \mathbf{x} \right]}_{i}}$ represents the the its $p$-norm and $i$-th element, respectively. For a general matrix $\bf{A}$, $\text{rank}(\bf{A})$, ${\bf{A}}^H$, ${{\mathbf{A}}^{\dagger}}$
 and ${{\left[\bf{A} \right]}_{i,j}}$ denote its rank, conjugate transpose, Moore-Penrose pseudo-inverse and $\left( i,j \right)$-th element, respectively. For a square matrix $\bf{X}$,  and $\text{tr}(\bf{X})$ denote its trace. ${\bf{X}} \succeq 0$ denotes that $\bf{X}$ is a positive semidefinite matrix. ${\mathbb{C}^{M \times N}}$ represents the ${M \times N}$ dimensional complex matrix space. $\mathbb{E}\left( \cdot  \right)$ denotes the expectation operation. $\sim$ represents “distributed as” and $\mathcal{C}\mathcal{N}\left( {\mathbf{x},\mathbf{R}} \right)$ represents
the distribution of a circularly symmetric complex Gaussian random vector with mean vector $\mathbf{x}$ and covariance matrix $\mathbf{R}$. $\otimes $ and $\odot$ represents the Kronecker and Hadamard product respectively.
\section{System Model and Problem Formulation}
As shown in Fig. 1, we consider a target-mounted IRS assisted-downlink ISAC system. The system consists of a multi-functional BS equipped with a uniform planar array (UPA) of $M$ active antennas, a single-antenna UE, and a mobile UAV target. To address the challenge of weak radar RCS typically associated with small UAVs, an IRS comprising $N$ passive reflective elements is mounted on the surface of the UAV. In this architecture, the IRS plays a critical dual role: it acts as a cooperative target to provide a strong LoS echo for self-tracking by the BS, and simultaneously functions as a mobile passive relay to establish a virtual LoS link, thereby enhancing the communication connectivity for the ground UE.

For tractability, a 3D Cartesian coordinate system is established. The BS is located at a fixed position, denoted by $\mathbf{p}_{\text{BS}}=[0,0,0]^{T}$. The UE is located at $\mathbf{p}_{\text{UE}}=[x_\text{U},y_\text{U},z_\text{U}]^{T}$, and its position is assumed to be known to the BS. The UAV carrying the IRS moves within the airspace. To describe its motion, we utilize the index $k$ to denote the $k$-th tracking epoch, a convention that applies throughout the remainder of this paper. We first define the position vector as $\mathbf{p}_{k}=[x_{k},y_{k},z_{k}]^{T}$ and the velocity vector as $\mathbf{v}_{k}=[\tilde{x}_{k},\tilde{y}_{k},\tilde{z}_{k}]^{T}$, where $\tilde{x}_{k}, \tilde{y}_{k}, \tilde{z}_{k}$ represent the instantaneous velocity components along the $x$, $y$, and $z$ axes, respectively. Consequently, the comprehensive state vector $\mathbf{x}_{k}$ is constructed by concatenating the position and velocity vectors, i.e.,
\begin{align}
\mathbf{x}_{k} = [\mathbf{p}_{k}^{T}, \mathbf{v}_{k}^{T}]^{T} = [x_{k}, y_{k}, z_{k}, \tilde{x}_{k}, \tilde{y}_{k}, \tilde{z}_{k}]^{T}.
\end{align}
The system operates on a discrete-time basis with a tracking interval of $\Delta T$. During the downlink transmission phase of the $k$-th epoch, the BS transmits a burst of $Q$ dual-functional pulses with a pulse repetition interval (PRI) of $T_p$ to achieve simultaneous communication and sensing. The duration of this pulse burst, referred to as the coherent processing interval (CPI), is $Q T_p$. To ensure the measurement process is completed within the current epoch, these parameters must satisfy the timing constraint $Q T_p \le \Delta T$. The system objectives are dual. First, for communication, the system aims to guarantee reliable data transmission to the UE via the direct link and the IRS-reflected link. Second, for sensing, the BS operates in a monostatic radar mode to estimate the UAV's state $\mathbf{x}_{k}$ in real-time. By processing the echo signals reflected by the IRS across the $Q$ pulses within the CPI, the BS can extract Doppler information and achieve high-precision tracking of the UAV's trajectory.
\begin{figure}%[htbp]
\centerline{\includegraphics[width=9cm]{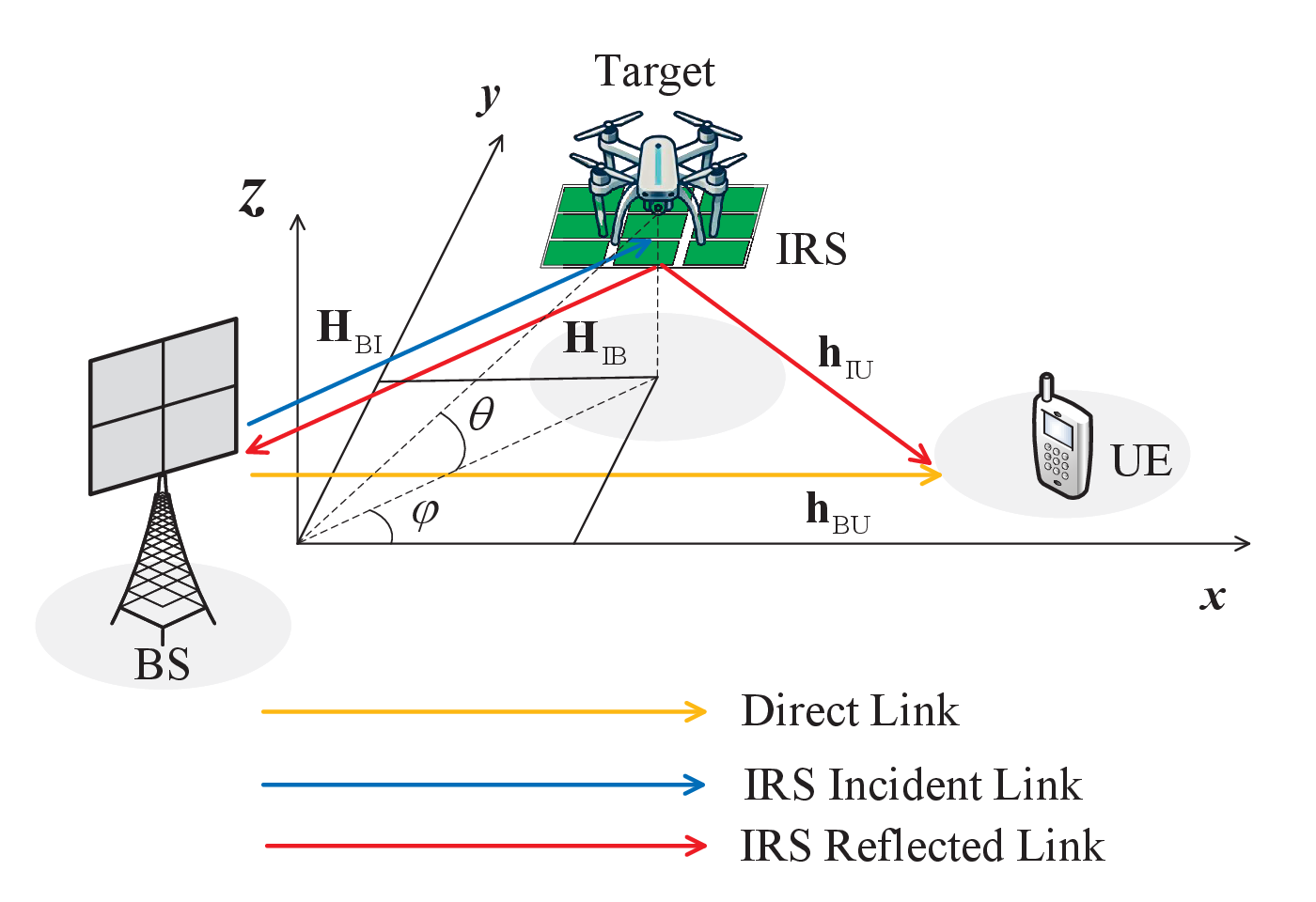}}
	\caption{Target-mounted IRS-assisted ISAC system.}
	\label{Fig1}
\end{figure}
\subsection{Channel Model}
In this section, we elaborate on the channel models for the considered target-mounted IRS-assisted ISAC system. Let $\mathbf{h}_{\text{BU},k} \in \mathbb{C}^{M \times 1}$, $\mathbf{h}_{\text{IU},k} \in \mathbb{C}^{N \times 1}$, and $\mathbf{H}_{\text{BI},k} \in \mathbb{C}^{N \times M}$ denote the channel vectors/matrix for the BS-UE direct link, the IRS-UE reflection link, and the BS-IRS link at the $k$-th time slot, respectively.

\subsubsection{Array Manifold and Steering Vectors}
We assume the BS is equipped with a UPA comprising $M = M_y \times M_z$ active antennas deployed in the $y-z$ plane, and the target-mounted IRS consists of a UPA with $N = N_x \times N_y$ passive reflecting elements arranged in the $x-y$ plane. For a signal propagating in a direction determined by the azimuth angle $\phi$ and elevation angle $\theta$, the array steering vector of the BS, denoted by $\mathbf{a}_{\text{BS}}(\phi, \theta) \in \mathbb{C}^{M \times 1}$, is given by
\begin{align}
\mathbf{a}_{\text{BS}}(\phi, \theta) = \mathbf{a}_{\text{BS}, z}(\theta) \otimes \mathbf{a}_{\text{BS}, y}(\phi, \theta),
\end{align}
where the vectors $\mathbf{a}_{\text{BS}, z}(\theta) \in \mathbb{C}^{M_z \times 1}$ and $\mathbf{a}_{\text{BS}, y}(\phi, \theta) \in \mathbb{C}^{M_y \times 1}$ represent the array response vectors along the $z$-axis and $y$-axis, respectively. Their $m$-th elements are defined as
\begin{align}
[\mathbf{a}_{\text{BS}, z}(\theta)]_m &= e^{-j {\pi}{\lambda}(m-1) \sin\theta}, \quad 1 \le m \le M_z, \\
[\mathbf{a}_{\text{BS}, y}(\phi, \theta)]_m &= e^{-j {\pi}{\lambda}(m-1) \cos\theta \sin\phi}, \quad 1 \le m \le M_y,
\end{align}
where a half-wavelength antenna spacing is assumed here and throughout the remainder of this paper. Similarly, the steering vector of the target-mounted IRS, $\mathbf{a}_{\text{IRS}}(\phi, \theta) \in \mathbb{C}^{N \times 1}$ can be expressed as
\begin{align}
\mathbf{a}_{\text{IRS}}(\phi, \theta) = \mathbf{a}_{\text{IRS}, y}(\phi, \theta) \otimes \mathbf{a}_{\text{IRS}, x}(\phi, \theta),
\end{align}
where $\mathbf{a}_{\text{IRS}, y}(\phi, \theta) \in \mathbb{C}^{N_y \times 1}$ and $\mathbf{a}_{\text{IRS}, x}(\phi, \theta) \in \mathbb{C}^{N_x \times 1}$ account for the phase shifts along the $y$-axis and $x$-axis, respectively. Their $n$-th elements are given by
\begin{align}
[\mathbf{a}_{\text{IRS}, y}(\phi, \theta)]_n &= e^{-j {\pi}{\lambda}(n-1) \cos\theta \sin\phi}, \quad 1 \le n \le N_y, \\
[\mathbf{a}_{\text{IRS}, x}(\phi, \theta)]_n &= e^{-j {\pi}{\lambda}(n-1) \cos\theta \cos\phi}, \quad 1 \le n \le N_x.
\end{align}

\subsubsection{Channel Links}
Based on the high-altitude deployment of the UAV, the channel links associated with the IRS are modeled as LoS channels, while the terrestrial BS-UE link is subject to multipath scattering. We adopt the free-space path loss model to characterize the signal attenuation. The channel matrix $\mathbf{H}_{\text{BI},k} \in \mathbb{C}^{N \times M}$ from the BS to the IRS is modeled as a rank-one LoS channel, strictly determined by the instantaneous geometric relationship between the fixed BS at $\mathbf{p}_{\text{BS}}$ and the mobile UAV at $\mathbf{p}_k$. Mathematically, this channel is expressed as
\begin{align}
\mathbf{H}_{\text{BI},k} = \sqrt{\beta_{\text{BI},k}} \mathbf{a}_{\text{IRS}}(\Omega_{\text{I},k}) \mathbf{a}_{\text{BS}}^H(\Omega_{\text{B},k}),
\end{align}
where $\beta_{\text{BI},k}$ represents the LoS path loss following the free-space propagation law. The notation $\Omega_{\text{B},k} = (\phi_k, \theta_k),$ denotes the angle of departure (\text{AoD}) from the BS, and $\Omega_{\text{I},k} = (\pi - \phi_k, \theta_k)$ denotes the angle of arrival (AoA) at the IRS, where $\phi_k$ and $\theta_k$ represent the azimuth and elevation angles in the global coordinate system at the $k$-th epoch. Similarly, the reflection channel vector $\mathbf{h}_{\text{IU},k} \in \mathbb{C}^{N \times 1}$ from the IRS to the UE is given by
\begin{align}
\mathbf{h}_{\text{IU},k} = \sqrt{\beta_{\text{IU},k}} \mathbf{a}_{\text{IRS}}(\Omega_{\text{IU},k}),
\end{align}
where $\beta_{\text{IU},k}$ is the path loss dependent on the distance $\|\mathbf{p}_k - \mathbf{p}_{\text{UE}}\|$, and $\Omega_{\text{IU},k}$ represents the \text{AoD} from the IRS towards the UE. In contrast, the direct link $\mathbf{h}_{\text{BU},k} \in \mathbb{C}^{M \times 1}$ from the BS to the ground UE is modeled as a Rician fading channel. It is composed of a deterministic LoS component $\mathbf{h}_{\text{BU},k}^{\text{LoS}}$ and a random NLoS Rayleigh fading component $\mathbf{h}_{\text{BU},k}^{\text{NLoS}} \sim \mathcal{CN}(\mathbf{0}, \mathbf{I}_M)$. The channel vector is formulated as
\begin{align}
\mathbf{h}_{\text{BU}} = \sqrt{\beta_{\text{BU}}} \left( \sqrt{\frac{\kappa}{\kappa+1}} \mathbf{h}_{\text{BU}}^{\text{LoS}} + \sqrt{\frac{1}{\kappa+1}} \mathbf{h}_{\text{BU}}^{\text{NLoS}} \right),
\end{align}
where $\kappa$ denotes the Rician factor representing the power ratio between the LoS and NLoS components and $\beta_{\text{BU}}$ represents the path loss. {\color{blue}While pure LoS links and perfect CSI are assumed for theoretical tractability, quantitative estimations reveal that the massive IRS aperture effectively hardens the channel. Under practical non-ideal conditions, such as Rician fading ($K=10$ dB) and imperfect CSI ($\sim 10^\circ$ phase error), the expected total array gain degradation is remarkably limited to merely $\sim 0.5$ dB, ensuring highly robust tracking and communication performance.}

\subsection{Signal Model}
In the $k$-th tracking epoch, the BS transmits a coherent pulse burst to simultaneously serve the downlink UE and track the target-mounted IRS. Let $s(t)$ denote the baseband waveform of a single pulse defined on $0 \le t \le T_p$ with unit energy ($\int_{0}^{T_p}|s(t)|^{2}dt=1$). The total transmitted baseband signal vector $\tilde{\mathbf{s}}_k(t) \in \mathbb{C}^{M \times 1}$ during the $k$-th epoch is modeled as a train of $Q$ pulses
\begin{align}\label{skt}
\tilde{\mathbf{s}}_k(t) = \mathbf{w}_k \sum_{q=0}^{Q-1} s(t - q T_p), \quad 0 \le t \le \Delta T,
\end{align}
where $\mathbf{w}_{k}\in\mathbb{C}^{M\times1}$ is the transmit beamforming vector, and the transmit power is $P_{t}=\|\mathbf{w}_{k}\|^{2}$.
\subsubsection{Communication Signal Model}
{\color{blue}The signal received at the UE from both direct link and the reflected link. Focusing on the signal received within the $q$-th pulse period (i.e., $t \in [q T_p, (q+1)T_p]$) and assuming perfect synchronization, the received signal $y_{u,k,q}(t)$ is
\begin{align}
y_{u,k,q}(t)\! =\! (\mathbf{h}_{\text{BU},k}^{H} \!+\! \mathbf{h}_{\text{IU},k}^{H}\mathbf{\Theta}_{k}\mathbf{H}_{\text{BI},k})\mathbf{w}_{k}s(t \!-\! q T_p)\! +\! n_{u,k}(t),
\end{align}}
where $n_{u,k}(t) \sim \mathcal{CN}(0, \sigma_u^2)$ denotes the additive white Gaussian noise (AWGN) at the user’s receiver
and $\mathbf{\Theta}_k = \mathrm{diag}(e^{j\psi_{k,1}}, \dots, e^{j\psi_{k,N}})$ represents the IRS phase shift matrix where $e^{j\psi_{k,n}}$ represents the reflection coefficient of the $n$-th element. Accordingly, the SNR of user $\gamma_{u,k}$ is given by 
\begin{align}
\gamma_{u,k} = \frac{|(\mathbf{h}_{\text{BU},k}^{H} + \mathbf{h}_{\text{IU},k}^{H}\mathbf{\Theta}_{k}\mathbf{H}_{\text{BI},k})\mathbf{w}_{k}|^{2}}{\sigma_{u}^{2}}. 
\end{align}
\subsubsection{Signal Model}
For sensing, the BS processes the echo signals reflected by the IRS. The signal travels along a round-trip path: BS $\to$ IRS $\to$ BS. First, we define the effective round-trip channel matrix $\mathbf{H}_{\text{rt},k} \in \mathbb{C}^{M \times M}$. Based on the channel reciprocity in time-division duplex  systems ($\mathbf{H}_{\text{IB},k} = \mathbf{H}_{\text{BI},k}^{T}$), it is given by the cascade
\begin{align}
\mathbf{H}_{\text{rt},k} \triangleq \mathbf{H}_{\text{IB},k} \mathbf{\Theta}_k \mathbf{H}_{\text{BI},k} = \mathbf{H}_{\text{BI},k}^{T} \mathbf{\Theta}_k \mathbf{H}_{\text{BI},k}.
\end{align}
The target's motion induces a round-trip delay $\tau_{k}=2\|\mathbf{p}_{k}\|/c$ and a Doppler frequency shift. Specifically, the Doppler shift $\nu_{k}$ is determined by the radial velocity $v_{\text{rad},k}$ of the UAV relative to the BS and can be expressed as
\begin{align}
\nu_{k} = \frac{2f_{c} v_{\text{rad},k}}{c} \quad \text{with} \quad v_{\text{rad},k} = \frac{\mathbf{p}_k^T \mathbf{v}_k}{\|\mathbf{p}_k\|}.
\end{align}
Based on the transmitted signal in (\ref{skt}), the received echo signal vector $\mathbf{r}_{k}(t)$ is a superposition of delayed and Doppler-shifted pulses. Considering the $q$-th pulse component ($q=0, \dots, Q-1$), the received signal $\mathbf{r}_{k,q}(t)$ corresponding to the transmitted term $\mathbf{w}_k s(t - q T_p)$ is modeled as
\begin{align}\label{r_kq}
\mathbf{r}_{k,q}(t) = \mathbf{H}_{\text{rt},k} \mathbf{w}_k s(t - q T_p - \tau_k) e^{j 2\pi \nu_k t} + \mathbf{n}_{s,k,q}(t), 
\end{align}
where $\mathbf{n}_{s,k,q}(t) \in \mathbb{C}^{M \times 1}$ represents the AWGN vector at the BS receive antenna array. Its entries are independent and identically distributed circularly symmetric complex gaussian (CSCG) random variables with zero mean and variance $\sigma_s^2$, denoted as $\mathbf{n}_{s,k,q}(t) \sim \mathcal{CN}(\mathbf{0}, \sigma_s^2 \mathbf{I}_M)$, By substituting the explicit expression of the round-trip channel $\mathbf{H}_{\text{rt},k}$ into (\ref{r_kq}) and isolating the spatial array gains, the noise-free echo signal is reformulated in terms of the continuous time $t$ as
\begin{align}
\tilde{\mathbf{r}}_{k,q}(t)\!=\!\beta_k \mathbf{a}_{\text{BS}}(\Omega_{\text{B},k}) s(t\!-\!qT_p\!-\!\tau_k) e^{-j2\pi f_c \tau_k} e^{j2\pi\nu_k t},
\end{align}
where $\mathbf{a}_{\text{BS}}(\Omega_{\text{B},k})$ is the receive array steering vector of the BS and
the scalar $\beta_k$ denotes the composite complex amplitude of the received echo. It compactly integrates the round-trip path loss, the IRS array gain, and the transmit array gain, given by
\begin{align}
\!\beta_k\!=\!\beta_{\text{BI},k}
\underbrace{\mathbf{a}_{\text{IRS}}^T(\Omega_{\text{I},k}) \mathbf{\Theta}_k \mathbf{a}_{\text{IRS}}(\Omega_{\text{I},k})}_{\text{IRS Array Gain}} 
\underbrace{\mathbf{a}_{\text{BS}}^H(\Omega_{\text{B},k}) \mathbf{w}_k}_{\text{Transmit Array Gain}},
\end{align}
where $e^{-j2\pi f_c \tau_k}$ represents the phase rotation caused by the propagation delay $\tau_k$ at the carrier frequency $f_c$, while $e^{j2\pi\nu_k t}$ accounts for the Doppler shift induced by the target motion.

\subsection{State Evolution Model}
In this section, we formulate the state space model for UAV tracking. We adopt the standard constant velocity model to describe the state evolution and establish a non-linear measurement model based on spatial frequencies.

\subsubsection{State Transition Model} 
Let $\Delta T$ denote the duration of one time slot. Recalling the definition in Section II-A, the UAV state vector is $\mathbf{x}_k = [x_k, y_k, z_k, \tilde{x}_k, \tilde{y}_k, \tilde{z}_k]^T$. The state evolution to the next time slot $k+1$ is modeled as
\begin{align}
\mathbf{x}_{k+1} = \mathbf{T} \mathbf{x}_k + \boldsymbol{\xi}_k,
\end{align}
where $\mathbf{T} = \mathbf{I}_3 \otimes \begin{bmatrix} 1 & \Delta T \\ 0 & 1 \end{bmatrix}$ is the state transition matrix. The process noise $\boldsymbol{\xi}_k \sim \mathcal{N}(\mathbf{0}, \mathbf{Q}_k)$ models the random accelerations. Specifically, for the CV model, the covariance matrix $\mathbf{Q}_k$ is given by
\begin{align}
\mathbf{Q}_k = q_s \mathbf{I}_3 \otimes \begin{bmatrix} \frac{\Delta T^3}{3} & \frac{\Delta T^2}{2} \\ \frac{\Delta T^2}{2} & \Delta T \end{bmatrix},
\end{align}
where $q_s$ denotes the power spectral density of the continuous-time process noise, reflecting the intensity of the UAV's motion.
\subsubsection{Observation Model} 
Upon receiving the echo signal, the BS extracts the key signal parameters to form the measurement vector $\mathbf{m}_k=[\tau_k,\nu_k,\mu_{y,k},\mu_{z,k} ] \in \mathbb{R}^{4 \times 1}$. To mitigate the linearization errors in the EKF, we employ spatial frequencies (denoted by $\mu$) instead of physical angles as the observable variables. The observation model is expressed as
\vspace{-0.2cm}
\begin{align}
\mathbf{m}_k = \boldsymbol{h}(\mathbf{x}_k) + \mathbf{n}_k,
\end{align}
where $\mathbf{n}_k \sim \mathcal{N}(\mathbf{0}, \mathbf{R}_k)$ denotes the measurement noise vector. The non-linear mapping $\boldsymbol{h}(\mathbf{x}_k)$ relates the state $\mathbf{x}_k$ to the signal parameters—namely the time delay $\tau_k$, Doppler shift $\nu_k$, and spatial frequencies along the $y$ and $z$ axes ($\mu_{y,k}, \mu_{z,k}$) and $\mathbf{h}(\mathbf{x}_{k})$ is given by
\vspace{-0.3cm}
\begin{align}\label{geo}
\boldsymbol{h}(\mathbf{x}_k) = \begin{bmatrix} 
\tau_k \\ 
\nu_k \\ 
\mu_{y,k} \\ 
\mu_{z,k} 
\end{bmatrix} 
= 
\begin{bmatrix} 
\frac{2 \|\mathbf{p}_k\|}{c} \\ 
\frac{2 f_c}{c} \frac{\mathbf{p}_k^T \mathbf{v}_k}{\|\mathbf{p}_k\|} \\ 
\frac{y_k}{\|\mathbf{p}_k\|} \\ 
\frac{z_k}{\|\mathbf{p}_k\|} 
\end{bmatrix}.
\end{align}
This formulation uses normalized coordinates $\mu_{y,k}$ and $\mu_{z,k}$ to avoid singularities in angle based measurements while facilitating the Jacobian matrix derivation for the tracking algorithm. It is important to note that the measurement noise covariance $\mathbf{R}_k$ is not constant because it fundamentally depends on the transmit signal power and the tracking geometry. Theoretically, $\mathbf{R}_k$ is lower bounded by the PCRB, the detailed derivation of which is presented in Section III.

\subsection{Problem Formulation} 
{\color{blue}Built upon the established signal and state models, this work jointly optimizes the active transmit beamforming vector $\mathbf{w}_k$ and the passive IRS phase-shift matrix $\mathbf{\Theta}_k$ to minimize total transmit power, subject to downlink communication quality of service (QoS) constraints and predefined UAV tracking accuracy requirements. UAV propulsion energy is omitted from the current formulation to prioritize transmission-layer sensing-communication co-optimization, and the holistic flight-communication energy efficiency design is reserved as our follow-up research.}

For the communication task, the received SNR $\gamma_{u,k}$ must exceed a minimum threshold $\gamma_{\text{com}}$. For the sensing task, we utilize the PCRB to characterize the fundamental limit of the estimation accuracy. Let $\mathbf{C}_k(\mathbf{w}_k, \mathbf{\Theta}_k) \triangleq \mathbf{F}_k^{-1}(\mathbf{w}_k, \mathbf{\Theta}_k)$ denote the predicted PCRB matrix derived from the Bayesian FIM $\mathbf{F}_k$. To ensure precise tracking, we impose separate constraints on the position and velocity estimation errors. By defining selection matrices $\mathbf{S}_{\text{pos}} = [\mathbf{I}_3, \mathbf{0}_{3 \times 3}]^T$ and $\mathbf{S}_{\text{vel}} = [\mathbf{0}_{3 \times 3}, \mathbf{I}_3]^T$, the bounds for the mean squared error (MSE) of position and velocity are extracted as $\text{Tr}(\mathbf{S}_{\text{pos}}^T \mathbf{C}_k \mathbf{S}_{\text{pos}})$ and $\text{Tr}(\mathbf{S}_{\text{vel}}^T \mathbf{C}_k \mathbf{S}_{\text{vel}})$, respectively. The explicit derivation of $\mathbf{F}_k$ will be presented in Section III-A. Mathematically, the joint optimization problem is formulated as
\vspace{-8pt}
\begin{align}
\mathcal{P}_1: \quad \min_{\mathbf{w}_k, \boldsymbol{\Theta}_k} \quad & \|\mathbf{w}_k\|^2 \\
\text{s.t.} \quad & \gamma_{u,k}(\mathbf{w}_k, \mathbf{\Theta}_k) \ge \gamma_{\text{com}}, \\
& \text{Tr}\left( \mathbf{S}_{\text{pos}}^T \mathbf{F}_k^{-1}(\mathbf{w}_k, \mathbf{\Theta}_k) \mathbf{S}_{\text{pos}} \right) \le \varepsilon_{\text{pos}}, \\
& \text{Tr}\left( \mathbf{S}_{\text{vel}}^T \mathbf{F}_k^{-1}(\mathbf{w}_k, \mathbf{\Theta}_k) \mathbf{S}_{\text{vel}} \right) \le \varepsilon_{\text{vel}}, \\
& |\psi_{k,n}| = 1, \quad \forall n = 1, \dots, N.
\end{align}
    The first constraint guarantees the communication rate. The second and third constraints enforce the accuracy requirements for the UAV's position and velocity, respectively, where $\varepsilon_{\text{pos}}$ and $\varepsilon_{\text{vel}}$ are the maximum tolerable MSE thresholds. The last constraint enforces the unit-modulus property of the IRS reflecting elements. $\mathcal{P}_1$ is a non-convex optimization problem due to the coupled variables in the SNR and PCRB terms, as well as the unit-modulus constraints. Efficient algorithms to solve this problem will be developed in Section IV.

\section{Target tracking framework and performance bound derivation}
{\color{blue}This section serves as a mathematical preliminary for the core research content, where the baseline target tracking framework and its fundamental performance limit are introduced with condensed formulation. We first briefly review the extended Kalman filter (EKF), which recursively estimates the target’s motion state by fusing prior motion predictions with radar measurements. We then derive the posterior Cramér–Rao bound (PCRB) to quantify the theoretical lower bound of 3D tracking accuracy. The original contributions of this work are detailed in the subsequent Section IV.}
\vspace{-12pt}
\subsection{Extended Kalman Filtering for State Estimation}
We employ the EKF to recursively estimate the dynamic UAV state $x_k$. The a priori state estimate $\hat{x}_{k|k-1}$ and error covariance $P_{k|k-1}$ are linearly predicted from the previous epoch:
\begin{align}
\hat{\mathbf{x}}_{k|k-1} &= \mathbf{T} \hat{\mathbf{x}}_{k-1|k-1}, \\
\mathbf{P}_{k|k-1} &= \mathbf{T} \mathbf{P}_{k-1|k-1} \mathbf{T}^T + \mathbf{Q}_k,
\end{align}
where $\mathbf{Q}_k$ represents the process noise covariance matrix at the $k$-th epoch. It accounts for the uncertainties in the state transition model. 
To incorporate the non-linear radar observation $\mathbf{y}_k$, we linearize the measurement model $\boldsymbol{h}(\mathbf{x}_k)$ around $\hat{\mathbf{x}}_{k|k-1}$ by computing the Jacobian matrix $\mathbf{H}_k \in \mathbb{R}^{4 \times 6}$:
\begin{align}
\mathbf{H}_k = \left. \frac{\partial \boldsymbol{h}(\mathbf{x})}{\partial \mathbf{x}} \right|_{\mathbf{x} = \hat{\mathbf{x}}_{k|k-1}} =
\begin{bmatrix}
\frac{\partial \tau}{\partial \mathbf{p}} & \mathbf{0}_{1 \times 3} \\
\frac{\partial \nu}{\partial \mathbf{p}} & \frac{\partial \nu}{\partial \mathbf{v}} \\
\frac{\partial \mu_y}{\partial \mathbf{p}} & \mathbf{0}_{1 \times 3} \\
\frac{\partial \mu_z}{\partial \mathbf{p}} & \mathbf{0}_{1 \times 3}
\end{bmatrix},
\end{align}
where the non-zero sub-blocks are analytically derived from the geometric relationships defined in (\ref{geo}). With the linearized measurement matrix, the Kalman gain is computed to minimize the a posterior error variance, i.e.,
\begin{align}
\mathbf{K}_k = \mathbf{P}_{k|k-1} \mathbf{H}_k^T \left( \mathbf{H}_k \mathbf{P}_{k|k-1} \mathbf{H}_k^T + \mathbf{R}_k \right)^{-1},
\end{align}
where $\mathbf{R}_k$ denotes the measurement noise covariance matrix derived in the subsequent section. Consequently, the a posterior state estimate is refined, and the estimation error covariance is updated to complete the recursive tracking loop, expressed as
\begin{align}
\hat{\mathbf{x}}_{k|k} &= \hat{\mathbf{x}}_{k|k-1} + \mathbf{K}_k \left( \mathbf{y}_k - \boldsymbol{h}(\hat{\mathbf{x}}_{k|k-1}) \right), \\
\mathbf{P}_{k|k} &= \left( \mathbf{I} - \mathbf{K}_k \mathbf{H}_k \right) \mathbf{P}_{k|k-1}.
\end{align}
\vspace{-10pt}
{\color{blue}\subsection{Derivation of the Posterior Cram\'er-Rao Bound}}
In this subsection, we derive the PCRB for the state vector $\mathbf{x}_{k}$ to quantify the theoretical performance limit of the tracking system. Let $\hat{\mathbf{x}}_{k|k}$ be any unbiased estimator based on the cumulative measurement history. Its MSE matrix is bounded from below by the inverse of the Bayesian FIM $\mathbf{F}_{k}$, denoted as $\mathbf{C}_{k}$. This fundamental relationship is expressed as
\begin{align}
\mathbb{E}[(\hat{\mathbf{x}}_{k|k}-\mathbf{x}_{k})(\hat{\mathbf{x}}_{k|k}-\mathbf{x}_{k})^{T}] \ge \mathbf{F}_{k}^{-1}.
\end{align}
The Bayesian FIM $\mathbf{F}_k$ represents the total information available at the current epoch. It consists of two parts: the predicted information $\mathbf{F}_{k|k-1}$ derived from the prior state evolution and the measurement information $\mathbf{F}_{\text{meas},k}$ from the current observation. The recursive update law is formulated as
\begin{align}
\mathbf{F}_k = \underbrace{(\mathbf{T}\mathbf{F}_{k-1}^{-1}\mathbf{T}^T + \mathbf{Q}_k)^{-1}}_{\mathbf{F}_{k|k-1}} + \underbrace{\mathbf{H}_k^T \mathbf{R}_k^{-1} \mathbf{H}_k}_{\mathbf{F}_{\text{meas},k}}
\end{align}
where $\mathbf{R}_k$ denotes the measurement noise covariance matrix. This matrix characterizes the estimation precision of the signal parameters embedded in the radar echo, specifically the time delay $\tau_k$, the Doppler shift $\nu_k$, and the spatial frequencies $\mu_{y,k}$ and $\mu_{z,k}$. By assuming parameter orthogonality within the high signal-to-noise ratio regime, we model $\mathbf{R}_k$ as a diagonal matrix, i.e.,
\vspace{-1mm}
\begin{align}
\mathbf{R}_{k} = \text{diag}(\sigma_{\tau,k}^{2}, \sigma_{\nu,k}^{2}, \sigma_{\mu_{y},k}^{2}, \sigma_{\mu_{z},k}^{2}).
\end{align}
The diagonal elements of $\mathbf{R}_{k}$ correspond to the CRB of the individual parameters. These are calculated based on the noise-free baseband signal model drived (\ref{r_kq}). Since the system transmits a burst of $Q$ coherent pulses, the Fisher information is accumulated over the pulse burst. For a specific parameter $\xi$, its variance is the inverse of the total Fisher information element $F_{\xi\xi}$, which can be calculated as
\begin{align}
\frac{1}{\sigma_{\xi,k}^{2}} = F_{\xi\xi} = \sum_{q=0}^{Q-1} \frac{2}{\sigma_{s}^{2}} \int_{0}^{T_p} \left\| \frac{\partial \tilde{\mathbf{r}}_{k,q}(t)}{\partial \xi} \right\|^{2} dt.
\end{align}
By solving this integral for each parameter, we obtain the closed-form expressions for the variances, explicitly incorporating the integration gain from $Q$ pulses.
\begin{align}
\sigma_{\tau,k}^{2} &= \frac{1}{8\pi^{2} Q \eta_{k} B_{\text{rms}}^{2}}, \quad \sigma_{\nu,k}^{2} = \frac{1}{8\pi^{2} Q \eta_{k} T_{\text{rms}}^{2}},\\\sigma_{\mu_{y},k}^{2} &= \frac{1}{8\pi^{2} Q \eta_{k} L_{y,\text{eff}}^{2}}, \quad \sigma_{\mu_{z},k}^{2} = \frac{1}{8\pi^{2} Q \eta_{k} L_{z,\text{eff}}^{2}}.
\end{align}
These results show that the tracking accuracy depends on the signal bandwidth $B_{\text{rms}}$, duration $T_{\text{rms}}$, the effective array apertures $L_{y,\text{eff}}^{2} \triangleq \frac{1}{M} \sum_{m=1}^{M} m_{y,m}^{2}, L_{z,\text{eff}}^{2} \triangleq \frac{1}{M} \sum_{m=1}^{M} m_{z,m}^{2}$, and is inversely proportional to the number of pulses $Q$. The sensing SNR is defined as $\eta_{k} \triangleq |\beta_{k}|^{2} M / \sigma_{s}^{2}$. Substituting the derived variance expressions into $\mathbf{R}_{k}$ fully defines the Bayesian FIM and the PCRB matrix.

{\color{blue}To fundamentally characterize the theoretical limits of the 3D tracking accuracy, we analytically quantify the exact contribution weights of the four observational dimensions to the state MSE. By extracting the $3 \times 3$ Jacobian sub-matrix $H_{pos}$ corresponding to the spatial measurements $[\tau_k, \mu_{y,k}, \mu_{z,k}]^T$, the instantaneous 3D position MSE is rigorously derived by computing the trace of the inverse FIM block
\begin{align}
&\text{MSE}_{\text{pos},k} = \text{Tr}\left( (\mathbf{H}_{\text{pos}}^T \mathbf{R}_{\text{pos}}^{-1} \mathbf{H}_{\text{pos}})^{-1} \right)\nonumber\\
&= \left(\frac{c}{2}\right)^2 \sigma_{\tau,k}^2 + \|p_k\|^2 \left( \frac{(1 - \mu_{z,k}^2)\sigma_{\mu_y,k}^2 + (1 - \mu_{y,k}^2)\sigma_{\mu_z,k}^2}{1 - \mu_{y,k}^2 - \mu_{z,k}^2} \right).
\end{align}

For 3D velocity tracking, the Doppler shift $\nu_k$ provides a critical rank-1 cross-correlation information matrix $\mathbf{F}_\nu = \sigma_{\nu,k}^{-2} \mathbf{H}_\nu^T \mathbf{H}_\nu$ to the global FIM. This strictly bounds the radial velocity error $\text{MSE}_{\text{vel},k}^{\text{radial}}$. Conversely, the cross-range velocity error $\text{MSE}_{\text{vel},k}^{\text{cross}}$ relies on the temporal fusion of spatial frequencies over the tracking interval $\Delta T$. Their theoretical bounds are respectively given by:
\begin{align}
&\text{MSE}_{\text{vel},k}^{\text{radial}} \ge \left(\frac{\lambda_c}{2}\right)^2 \sigma_{\nu,k}^2, \\
&\text{MSE}_{\text{vel},k}^{\text{cross}} \propto \frac{\|p_k\|^2}{\Delta T^2} \left( \frac{(1 - \mu_{z,k}^2)\sigma_{\mu_y,k}^2 + (1 - \mu_{y,k}^2)\sigma_{\mu_z,k}^2}{1 - \mu_{y,k}^2 - \mu_{z,k}^2} \right).
\end{align}}

\section{Performance Trade-off Analysis and Efficient Algorithm Design}
Due to the high mobility of the UAV, traditional complex algorithms (e.g., SDR) cause high processing delay, which fails to meet the requirement for real-time tracking. To address this, we analyze the fundamental trade-off between sensing and communication. By using the geometric structure of the problem, we propose a fast algorithm with very low computational complexity.

\subsection{Transformation of Performance Constraints into Amplitude Thresholds}
The joint optimization problem has different requirements for sensing accuracy and communication quality, which involves complex constraints. To make the problem easier to solve, we convert these performance requirements into clear signal amplitude thresholds. This step significantly simplifies the problem structure and helps to design an efficient algorithm.

\subsubsection{Derivation of Sensing Amplitude Threshold}
The sensing performance depends on the magnitude of the noise-free echo received at the BS. The sensing amplitude, denoted as $A_{s,k}$, can be expressed as
\begin{align}
A_{s,k}(\mathbf{w}_k, \mathbf{\Theta}_k) =\left| \sqrt{\beta_{\text{BI},k}} \mathbf{a}_{\text{IRS}}^T \mathbf{\Theta}_k \mathbf{a}_{\text{IRS}} \mathbf{a}_{\text{BS}}^H \mathbf{w}_k \right|. 
\end{align}
The theoretical lower bounds for the UAV’s position and velocity estimation errors are determined by the trace of the inverse $\mathbf{F}_k$. By utilizing the definition of the sensing amplitude $A_{s,k}$, $\mathbf{F}_k$ can be reformulated as
\begin{align}
\mathbf{F}_k(A_{s,k}) = \mathbf{F}_{k|k-1} + \frac{M A_{s,k}^2}{\sigma_s^2} \mathbf{\Psi}_k,
\end{align}
where $\mathbf{\Psi}_k$ represents the normalized observation information matrix, which captures the geometric contribution from the measurement noise covariance $\mathbf{R}_k$.

The sensing constraints mandate that the predicted MSE for position and velocity do not exceed the pre-defined thresholds $\epsilon_{\text{pos}}$ and $\epsilon_{\text{vel}}$. Let $g_{\text{pos}}(A_{s,k}) = \text{Tr}(\mathbf{S}_{\text{pos}}^T \mathbf{F}_k^{-1} \mathbf{S}_\text{{pos}})$ and $g_\text{{vel}}(A_{s,k}) = \text{Tr}(\mathbf{S}_{vel}^T \mathbf{F}_k^{-1} \mathbf{S}_{\text{vel}})$ denote the MSE functions for position and velocity, respectively. 

Although deriving a closed-form expression for $A_{s,k}$ is mathematically intractable due to the matrix inversion, the positive semi-definite nature of $\Psi_k$ ensures that a larger $A_{s,k}$ strictly increases the Fisher information $\mathbf{F}_k$, thereby reducing estimation uncertainty. Consequently, the MSE functions $g_{\text{pos}}(A_{s,k})$ and $g_{\text{vel}}(A_{s,k})$ are strictly monotonically decreasing with respect to $A_{s,k}$. This monotonicity mathematically guarantees unique critical amplitudes $A_{\text{pos},k}$ and $A_{\text{vel},k}$ that satisfy the equality constraints, efficiently determined via bisection.To satisfy both accuracy constraints simultaneously, the effective sensing amplitude threshold, denoted as 
\begin{align}
A_{s,k}\geq\Gamma_{s,k} = \max\{A_{\text{pos},k}, A_{\text{vel},k}\}.
\end{align}

\subsubsection{Derivation of Communication Amplitude Threshold} The communication quality is determined by the received signal amplitude at the UE and is denoted as $A_{c,k}$
\begin{align}
A_{c,k}(\mathbf{w}_k, \mathbf{\Theta}_k) \triangleq \left| (\mathbf{h}_{\text{BU},k}^H + \mathbf{h}_{\text{IU},k}^H \mathbf{\Theta}_k \mathbf{H}_{\text{BI},k}) \mathbf{w}_k \right|. 
\end{align}
Unlike the sensing metric, the communication constraint admits a closed-form solution. By enforcing the QoS requirement $\text{SNR} \ge \gamma_{\text{req}}$, the minimum required amplitude is directly obtained from the noise power $\sigma_u^2$. Thus, the communication amplitude threshold $\Gamma_{c,k}$ is derived as
\begin{align}
A_{c,k}\geq\Gamma_{c,k} = \sigma_u \sqrt{\gamma_{\text{req}}}.
\end{align}
\subsection{Geometric Analysis of Performance Trade-off: The Elliptical Boundary}
To analyze the fundamental trade-off between the sensing amplitude $A_{s,k}$ and the communication amplitude $A_{c,k}$, we perform a geometric analysis based on the channel vectors. The original optimization problem involves high-dimensional variables, which makes the trade-off relationship difficult to observe. To derive the analytical boundary, we apply the subspace projection method to reduce the problem dimension.

The performance limit of the ISAC system depends on the spatial separation between the sensing target and the communication user. To quantify this geometric limit, we define the spatial correlation metrics for both the active and passive arrays. At the BS side, the spatial correlation between the direct communication channel $\mathbf{h}_{\text{BU},k}$ and the IRS LoS channel $\mathbf{a}_{\text{BS}}$ is quantified by the angle $\rho_{\text{BS}}$, which can be expressed as
\begin{align}
\rho_{\text{BS},k} = \arccos\left( \frac{|\mathbf{h}_{\text{BU}}^H \mathbf{a}_{\text{BS}}(\Omega_{\text{B},k})|}{\|\mathbf{h}_{\text{BU}}\| \|\mathbf{a}_{\text{BS}}(\Omega_{\text{B},k})\|} \right), 
\end{align} 
where $\rho_{\text{BS},k} \in [0, \pi]$. Physically, a smaller $\rho_{\text{BS}}$ indicates a higher spatial correlation. Specifically, $\rho_{\text{BS},k} = 0$ implies that the two channels are spatially aligned. In this ideal case, the integrated waveform can simultaneously achieve the full beamforming gain for both functions without power splitting.

Similarly, at the IRS side, the signal transmission depends on the cascaded channel gain, which combines the incident and reflected steering vectors through the IRS phase shifts. To linearize this relationship, we define the effective spatial vector $\mathbf{d}$ as the Hadamard product of the input and output steering vectors. Specifically, the effective sensing vector is denoted as $\mathbf{d}_{\text{\text{sen}},k} = \mathbf{a}_{\text{IRS}}(\Omega_{\text{I},k}) \odot \mathbf{a}_{\text{IRS}}(\Omega_{\text{I},k})$ and the effective communication vector as $\mathbf{d}_{_\text{com}} = \mathbf{a}_{\text{IRS}}(\Omega_{\text{I},k}) \odot \mathbf{a}_{\text{IRS}}^*(\Omega_{\text{IU},k})$. The spatial correlation between these two vectors is quantified by the angle $\rho_{\text{IRS},k}$, which can be expressed as
\begin{align}
\rho_{\text{IRS},k} = \arccos\left( \frac{|\mathbf{d}_{\text{sen},k}^H \mathbf{d}_{\text{com},k}|}{\|\mathbf{d}_{\text{sen},k}\| \|\mathbf{d}_{\text{com},k}\|} \right),
\end{align}
where $\rho_{\text{IRS},k} \in [0, \pi]$. Physically, a smaller $\rho_{\text{IRS},k}$ indicates a higher correlation between the two functional requirements. Specifically, $\rho_{\text{IRS},k} = 0$ implies that the optimal phase configurations for sensing and communication are identical. In this ideal case, the IRS can simultaneously maximize the cascade channel gain for both links without any trade-off.  

{\color{blue}Geometrically, the scalar parameters $\lambda_{\text{BS},k}$ and $\lambda_{\text{IRS},k}$ physically define the energy projection ratios of the active and passive beams between the decoupled sensing and communication subspaces. Specifically, the projection amplitudes of the active BS beam onto the sensing and communication directions are quantified by $\cos(\lambda_{\text{BS},k})$ and $\cos(\rho_{\text{BS},k} - \lambda_{\text{BS},k})$, respectively. Similarly, $\lambda_{\text{IRS},k}$ regulates the passive reflection power distribution via $\cos(\lambda_{\text{IRS},k})$ for radar backscattering and $\cos(\rho_{\text{IRS},k} - \lambda_{\text{IRS},k})$ for forward user relaying. Consequently, optimizing these scalars dynamically steers the spatial power distribution to match the instantaneous tracking and QoS requirements. }

Based on the derived spatial correlation, the complex high-dimensional optimization for the active beamforming vector $\mathbf{w}_k$ and the passive IRS phase shifts $\mathbf{\Theta}_k$ can be significantly simplified. Since the optimal signal transmission must be concentrated in the subspaces spanned by the respective channel vectors, the search for the optimal strategy reduces to two scalar angular parameters: $\lambda_{\text{BS},k} \in [0, \rho_{\text{BS},k}]$ for the BS active beamforming and $\lambda_{\text{IRS},k} \in [0, \rho_{\text{IRS},k}]$ for the IRS passive beamforming. These scalar parameters represent the angular deviations from the respective channel references. To avoid unnecessary power leakage, the effective steering directions must be confined within the angular region bounded by the functional vectors. Any beam direction falling outside the interval $[0, \rho_k]$ would result in the beam being steered away from both the sensing and communication directions simultaneously. Therefore, $\mathbf{w}_k$ and $\mathbf{\Theta}_k$ can be expressed as
\begin{align}
\mathbf{w}_k = \sqrt{P} \frac{\lambda_{\text{BS},k} \mathbf{a}_{\text{BS}}(\Omega_{\text{B},k}) + (1 - \lambda_{\text{BS},k}) \mathbf{h}_{\text{BU}}}{\|\lambda_{\text{BS},k} \mathbf{a}_{\text{BS}}(\Omega_{\text{B},k}) + (1 - \lambda_{\text{BS},k}) \mathbf{h}_{\text{BU}}\|},
\end{align}
and $\mathbf{\Theta}_k = \text{diag}(\boldsymbol{\theta}_k)$ with
\begin{align}
\boldsymbol{\theta}_k = \lambda_{\text{IRS},k} \mathbf{a}_{\text{IRS}}^{*}(\Omega_{\text{I},k}) + (1 - \lambda_{\text{IRS},k}) {\mathbf{a}_{\text{IRS}}(\Omega_{\text{IU},k})}.
\end{align}
Consequently, we can parameterize the system performance by using $\lambda_{\text{BS},k}$ and $\lambda_{\text{IRS},k}$. For a fixed active beam direction $\lambda_{\text{BS},k}$, the effective gain coefficients for the direct communication path $C_{\text{dir},k}$, the reflected communication path $C_{\text{ref},k}$, and the sensing path $C_{\text{sen},k}$ can be expressed as
\begin{align}
C_{\text{dir},k}& = \sqrt{P_t} \|\mathbf{h}_{\text{BU}}\| \cos(\rho_{\text{BS},k} - \lambda_{\text{BS},k}), \\
C_{\text{ref},k} &= \sqrt{P_t} |\beta_{\text{BI},k} \beta_{\text{IU},k}| MN \cos(\lambda_{\text{BS},k}),\\ 
C_{\text{sen},k}& = \sqrt{P_t} |\beta_{\text{BI},k}|^2 MN \cos(\lambda_{\text{BS},k}).
\end{align}

With the gain coefficients $C_{\text{dir},k}$, $C_{\text{ref},k}$, and $C_{\text{sen},k}$ determined by the active angle $\lambda_{\text{BS},k}$, the final signal amplitudes are formulated as functions of both $\lambda_{\text{BS},k}$ and $\lambda_{\text{IRS},k}$. By combining the active and passive beamforming effects, the communication and sensing amplitudes are derived as
\begin{align}\label{A_ck}
A_{c,k}(\lambda_{\text{BS},k}, \lambda_{\text{IRS}}) = C_{\text{dir},k} + C_{\text{ref},k} \cos(\rho_{\text{IRS},k} - \lambda_{\text{IRS},k}), 
\end{align}
and
\begin{align}\label{A_sk}
A_{s,k}(\lambda_{\text{BS},k}, \lambda_{\text{IRS},k}) = C_{\text{sen},k} \cos(\lambda_{\text{IRS},k}). 
\end{align}
To characterize the boundary of the achievable region, we eliminate the passive steering angle $\lambda_{\text{IRS},k}$ to establish a direct relationship between the sensing and communication amplitudes. The derivation proceeds in three steps. First, from (\ref{A_sk}), the cosine component is directly determined as
\begin{align}
\cos(\lambda_{\text{IRS},k}) = \frac{A_{s,k}}{C_{\text{sen},k}}. 
\end{align}
Next, we substitute this ratio into (\ref{A_ck}) and apply the trigonometric expansion formula to $\cos(\rho_{\text{IRS},k} - \lambda_{\text{IRS},k})$. By isolating the term containing $\sin(\lambda_{\text{IRS},k})$, we obtain:
\begin{align}\label{elli1}
\frac{A_{c,k} - C_{\text{dir},k}}{C_{\text{ref},k}} - \frac{A_{s,k}}{C_{\text{sen},k}}\cos(\rho_{\text{IRS},k}) = \sin(\rho_{\text{IRS},k})\sin(\lambda_{\text{IRS},k}). 
\end{align}
Finally, we square both sides of (\ref{elli1}) and utilize the fundamental identity $\sin^2(\lambda_{\text{IRS},k}) = 1 - \cos^2(\lambda_{\text{IRS},k})$. After rearranging the terms to group the amplitude variables $A_{c,k}$ and $A_{s,k}$, we can obtain a general elliptical equation, which can be expressed as (\ref{insights}).
\begin{figure*}[!t]
	\normalsize
\begin{align}\label{insights}
\frac{(A_{c,k} - C_{\text{dir},k})^2}{C_{\text{ref},k}^2} + \frac{A_{s,k}^2}{C_{\text{sen},k}^2} - \frac{2(A_{c,k} - C_{\text{dir},k})A_{s,k} \cos(\rho_{\text{IRS},k})}{C_{\text{ref},k} C_{\text{sen},k}} = \sin^2(\rho_{\text{IRS},k}).
\end{align}
\hrulefill
\vspace*{1pt}
\end{figure*}

\subsection{Optimal Solution Strategy: A Low-Complexity Algorithm based on Case Analysis}
To efficiently determine the minimum transmit power $P_t^*$ and the corresponding optimal joint beamforming strategy $(\lambda_{\text{BS},k}^*, \lambda_{\text{IRS},k}^*)$ that satisfy the QoS requirements $\Gamma_{s,k}$ and $\Gamma_{c,k}$, we propose a low-complexity algorithm based on case analysis. By exploiting the geometric property that the optimal solution must reside on the boundary of the feasible region, the continuous optimization problem is transformed into a discrete selection among potential candidates. This approach allows us to secure the global optimum in closed form without the need for computationally intensive iterative searches.

To facilitate the mathematical derivation, we decouple the transmit power from the geometric gains by defining the unit-power effective gain coefficients, denoted as $\bar{C}$. These coefficients quantify the intrinsic channel strengths achievable with a unit transmit power ($P_t = 1$ W) for a given active steering angle $\lambda_{\text{BS}}$, i.e.,
\begin{align}
\bar{C}_{\text{dir},k}(\lambda_{\text{BS},k}) &= \|\mathbf{h}_{\text{BU}}\| \cos(\rho_{\text{BS},k} - \lambda_{\text{BS},k}), \\
\bar{C}_{\text{ref},k}(\lambda_{\text{BS},k}) &= \|\mathbf{h}_{\text{BI},k}\mathbf{h}_{\text{IU},k}\| MN \cos(\lambda_{\text{BS},k}), \\
\bar{C}_{\text{sen},k}(\lambda_{\text{BS},k}) &= \|\mathbf{h}_{\text{BI},k}\|^2 MN \cos(\lambda_{\text{BS},k}).
\end{align} 
Based on these normalized coefficients, the optimization problem can be analytically solved by examining three distinct cases. These cases are classified by comparing the required amplitude thresholds imposed by the sensing and communication tasks.

\subsubsection{Sensing-Dominant Regime ($A_{s,k} = \Gamma_{s,k},A_{c,k} \ge \Gamma_{c,k}$)} In this case, the system performance is strictly limited by the sensing requirement. To satisfy the sensing threshold with minimal power consumption, the optimal strategy is to maximize the effective sensing gain. Consequently, both the active and passive beams are perfectly aligned with the sensing channel directions, i.e.,
\begin{align}
\lambda_{\text{BS},k}^{*} = 0, \quad \lambda_{\text{IRS},k}^{*} = 0.
\end{align}
With this configuration, the minimum transmit power required to meet the binding sensing constraint is derived by inverting the sensing gain function, which can be expressed as
\begin{align}
P^{(1)}_k = \left( \frac{\Gamma_{s,k}}{\bar{C}_{\text{sen},k}(0)} \right)^{2}.
\end{align}
This candidate solution is valid only if the communication constraint is naturally satisfied. We verify this by checking whether the resulting communication amplitude exceeds $\Gamma_{c,k}$. If the condition fails, we assign an infinite penalty to this candidate, which is formulated as
\begin{align}
\!P_{\text{cand},k}^{(1)}\!=\! \begin{cases} P^{(1)}_k\!,\!\text{if }\!\sqrt{P^{(1)}_k} \left( \bar{C}_{\text{dir},k} + \bar{C}_{\text{ref},k}\cos(\rho_{\text{IRS},k}) \right)\! \ge\! \Gamma_{c,k} \\ \infty,\! \text{otherwise}.\end{cases}
\end{align}
\subsubsection{Communication-Dominant Regime ($A_{c,k} = \Gamma_{c,k}, A_{s,k} \ge \Gamma_{s,k}$)} In this case, the system prioritizes the communication link. To maximize the received signal strength at the user, the IRS is configured to maximize the specular reflection, while the active beam is optimized via the maximum ratio transmission principle to achieve the coherent combination of the direct and reflected paths. Accordingly, the optimal steering angles are analytically determined as
\begin{align}
\lambda_{\text{IRS},k}^{*} = \rho_{\text{IRS},k},
\end{align}
and
\begin{align}
\lambda_{\text{BS},k}^{*}\!=\!\arctan \left( \frac{\|\mathbf{h}_{\text{BU},k}\|\sin(\rho_{\text{BS},k})}{\|\mathbf{h}_{\text{BU}}\|\cos(\rho_{\text{BS,k}}) + \|\beta_{\text{BU}}\beta_{\text{BU}}\|MN} \right).
\end{align}
\vspace{-1mm}
Based on these optimal angles, the minimum transmit power required to satisfy the binding communication threshold is computed by inverting the effective communication gain, which is given by
\begin{align}
P^{(2)}_k = \left( \frac{\Gamma_{c,k}}{\bar{C}_{\text{dir},k}(\lambda_{\text{BS},k}^{*}) + \bar{C}_{\text{ref},k}(\lambda_{\text{BS},k}^{*})} \right)^{2}.
\end{align}
Similar to case 1, we validate this solution by checking the sensing performance. If the resulting sensing amplitude meets $\Gamma_{s,k}$, the solution is valid; otherwise, it is marked as infeasible, which implies
\begin{align}
P_{\text{cand},k}^{(2)}\!=\!\begin{cases} P^{(2)}_k\!,\! & \text{if } \sqrt{P^{(2)}_k} \bar{C}_{\text{sen},k}(\lambda_{\text{BS},k}^{*})\cos(\rho_{\text{IRS},k}) \ge \Gamma_{s,k} \\ \infty,\! & \text{otherwise}. \end{cases}
\end{align}

\subsubsection{Coupled Constrained Case ($A_{s,k} = \Gamma_{s,k}, A_{c,k} = \Gamma_{c,k}$)} This case corresponds to the critical Pareto frontier where both constraints are strictly active. By substituting the fixed amplitudes $A_{s,k}=\Gamma_{s,k}$ and $A_{c,k}=\Gamma_{c,k}$ into the elliptical trade-off equation derived previously, the problem simplifies to solving for the unknown power scaling factor $x=\sqrt{P_t}$. This algebraic manipulation yields a standard univariate quadratic equation, which is written as
\begin{align}
\mathcal{A}x^{2} + \mathcal{B}x + \mathcal{C} = 0.
\end{align}
where the coefficients $\mathcal{A}$, $\mathcal{B}$ and $\mathcal{C}$ are functions of the active beam angle $\lambda_{\text{BS}}$ and are defined as
\begin{align}
\mathcal{A} &= \bar{C}_{\text{dir},k}^2 + \bar{C}_{\text{ref},k}^2\sin^2(\rho_{\text{IRS},k}), \\
\mathcal{B} &= -2 \Gamma_c \bar{C}_{\text{dir},k}(\lambda_{\text{BS},k}) - 2 \Gamma_{s,k} \frac{\bar{C}_{\text{dir},k}\bar{C}_{\text{ref},k}}{\bar{C}_{\text{sen},k}} \cos(\rho_{\text{IRS}}), \\
\mathcal{C} &= \Gamma_{c,k}^2 + \left( \frac{\bar{C}_{\text{ref},k}}{\bar{C}_{\text{sen},k}} \Gamma_{s,k} \right)^2 - \frac{2 \Gamma_c \Gamma_s \bar{C}_{\text{ref},k}}{\bar{C}_{\text{sen},k}} \cos(\rho_{\text{IRS},k}).
\end{align}
For any given $\lambda_{\text{BS},k}$, the required power $P_{\text{req}}(\lambda_{\text{BS},k})$ corresponds to the square of the positive root of the quadratic equation. Accordingly, the optimal active beam angle is determined by minimizing this power using a low-complexity 1D bisection search, which implies
\begin{align}
\lambda_{\text{BS},k}^{*} = \arg \min_{\lambda \in [0, \rho_{\text{BS},k}]} P_{\text{req}}(\lambda), \quad P_{\text{cand},k}^{(3)} = P_{\text{req}}(\lambda_{\text{BS},k}^{*}).
\end{align}
Finally, the corresponding optimal IRS angle is uniquely recovered from the active sensing constraint, formulated as
\begin{align}
\lambda_{\text{IRS},k}^{*} = \arccos \left( \frac{\Gamma_{s,k}}{\sqrt{P_{\text{cand},k}^{(3)}}  \bar{C}_{\text{sen},k}(\lambda_{\text{BS},k}^{*})} \right).
\end{align}
\vspace{-20pt}
\subsection{Global Optimal Solution Selection}
Upon deriving the candidate solutions for the three distinct cases, the global optimal transmit power is determined by selecting the strategy that yields the minimum power consumption among the valid candidates. Since any infeasible candidate has been assigned an infinite penalty in the previous steps, the final optimal transmit power $P_{t,k}^*$ is simply obtained as
\begin{align}
P_{t,k}^{*} = \min \{ P_{\text{cand},k}^{(1)}, P_{\text{cand},k}^{(2)}, P_{\text{cand},k}^{(3)} \}.
\end{align}
Accordingly, the optimal joint beamforming strategy $(\lambda_{\text{BS},k}^*, \lambda_{\text{IRS},k}^*)$ is identified as the configuration associated with the selected minimum power candidate. This selection process guarantees that the system operates at the global optimum, satisfying both the sensing and communication requirements with minimum power consumption.

\section{Simulation Results}
We simulate the proposed ISAC system in a 3D Cartesian coordinate system. BS is fixed at $\textbf{p}_{\text{BS}}=[0, 0, 30]^T$ m, and the UAV equipped with the IRS initializes at $\textbf{p}_{\text{IRS}}=[30, 30, 50]^T$ m, with UE positions varying per geometric scenario. Key simulation parameters are summarized in Table I.

To verify the effectiveness of the proposed joint algorithm, we compare it against four benchmark schemes under identical conditions : 1) No-IRS: Relies solely on direct links for communication and sensing without IRS assistance; 2) Random-IRS: Utilizes random IRS phase shifts, offering passive aperture gain without directional beamforming; 3) TDM: Orthogonally switches between optimal sensing and communication beams in the time domain ; and 4) Align-to-UE: Strictly aligns the IRS reflected beam toward the UE to maximize communication signal strength.

{\color{blue}It is mathematically crucial to emphasize that the performance curve of the proposed scheme explicitly represents the exact theoretical Pareto frontier derived via strict geometric analysis. Unlike conventional state-of-the-art numerical algorithms that asymptotically search for local optima and suffer from mathematical relaxation gaps, the proposed analytical solution directly yields the exact closed-form global optimum. Consequently, it inherently constitutes the absolute theoretical performance ceiling of the target-mounted ISAC architecture, effectively superseding the need for empirical comparisons with numerical approximation solvers. The selected benchmark schemes (i.e., No-IRS, Random-IRS, TDM, and Align-to-UE) instead serve to physically validate the architectural superiority of the target-mounted mechanism and the fundamental advantages of coherent spatial multiplexing over orthogonal resource allocation.}

{\color{blue}
\begin{table}[t]
\centering
\captionsetup{justification=centering, singlelinecheck=true}
\small
\caption{Simulation Parameters}
\scalebox{0.88}{
\begin{tabular}{|c|c|c|c|c|c|}
\hline Parameter & Value & Parameter & Value & Parameter & Value\\
\hline$M$ & $64$ & ${N}$ & $64$ & $T_p$ & $10^{-3}$\\
\hline$W$ & $500\text{MHz}$ & ${K}$ & $100$ & $\kappa $ & $7 \text{dBsm}$\\
\hline${{P}_{\max}}$ & $30\text{dBm}$ & $\sigma^{2}$ & $-90\text{dBm}$ & $\lambda$ & $0.1\text{m}$\\
\hline
\end{tabular}}
\end{table}}
\subsection{Performance Trade-off Analysis}
In this subsection, we investigate the fundamental performance limits of the ISAC system by analyzing the Pareto Optimality between sensing and communication amplitudes. The Pareto frontier characterizes the upper bound of performance for sensing and communication functions under given power and hardware constraints. Within this theoretical framework, we first validate the effectiveness of the proposed joint beamforming design in approaching the Pareto optimality by comparing it against benchmark schemes. Furthermore, we analyze the expansion effect of the number of IRS reflecting elements on the Pareto frontier and investigate the evolution and robustness of this frontier under various spatial geometric configurations.
\subsubsection{Pareto Frontier of Sensing and Communication Versus Benchmark Schemes}
\begin{figure}%[htbp]
\centerline{\includegraphics[width=1\columnwidth]{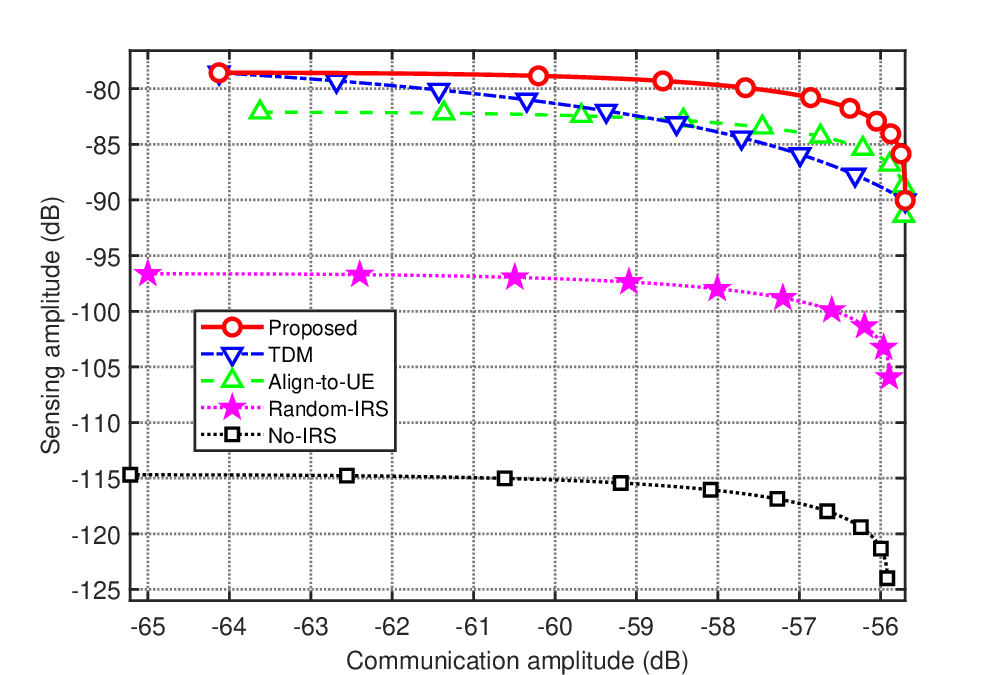}}
	\caption{Pareto frontier of sensing and communication with different
optimization schemes.}\label{p1}
\end{figure}
Fig. 2 illustrates the sensing-communication performance trade-off of the proposed joint beamforming scheme versus benchmark schemes where the UE is located at the coordinate [20, 20, 2] meters. The proposed scheme characterizes the optimal boundary of sensing and communication amplitudes, yielding a significantly larger achievable region compared to all benchmark schemes. {\color{blue}Specifically, the proposed scheme consistently outperforms the TDM baseline by achieving coherent waveform superposition rather than orthogonal time-resource partitioning, fully exploiting spatial degrees of freedom. 

Furthermore, while the ``Align-to-UE" scheme matches the proposed optimal communication amplitude, it suffers a severe sensing loss exceeding 15 dB because strictly aligning the reflected beam toward the UE causes massive deviation from the radar target direction. This validates that rigid single-objective beamforming fails in complex dual-functional environments.}

{\color{blue} Analytically, the total sensing power enhancement provided by the target-mounted IRS scales with $N^2$. This total gain can be mathematically decoupled into two orthogonal components: an $N$-fold passive aperture gain stemming from non-coherent physical scattering (represented by the ``Random-IRS" benchmark, yielding $\mathbb{E}[|\sum e^{j\phi}|^2] = N$), and an additional $N$-fold directional beamforming gain achieved via coherent phase alignment. For our $N=64$ configuration, this translates to a symmetric logarithmic contribution: 18 dB from the passive aperture and 18 dB from active beamforming, cumulatively providing a ~36 dB power margin to reliably overcome the UAV's inherently weak radar cross-section.}

However, regarding communication performance, the reflected signals with random phases fail to form a directional beam towards the user. It is crucial to note that the marginal performance gain in this specific case is also attributed to the geometric proximity between the UE and the BS. Since the user is located relatively close to the transmitter, the strong line-of-sight direct link dominates the communication channel capacity, rendering the contribution of the passive reflection link less significant compared to the sensing scenario. The detailed impact of varying user locations and distances on the system performance will be explicitly analyzed in the subsequent subsection.

\subsubsection{Pareto Frontier of Sensing and Communication Versus Number of Reflecting Elements}
\begin{figure}[htbp]
    \centering
            \captionsetup[subfigure]{justification=centering}
    % --- 第一个子图 (a) ---
    % 宽度设为 1.0\textwidth 表示占满一行，这样第二个图就会自动换行到下面
    \begin{subfigure}[b]{\columnwidth}
        \centering
        % 插入图片，这里 width 控制图片实际大小
        \includegraphics[width=0.95\columnwidth]{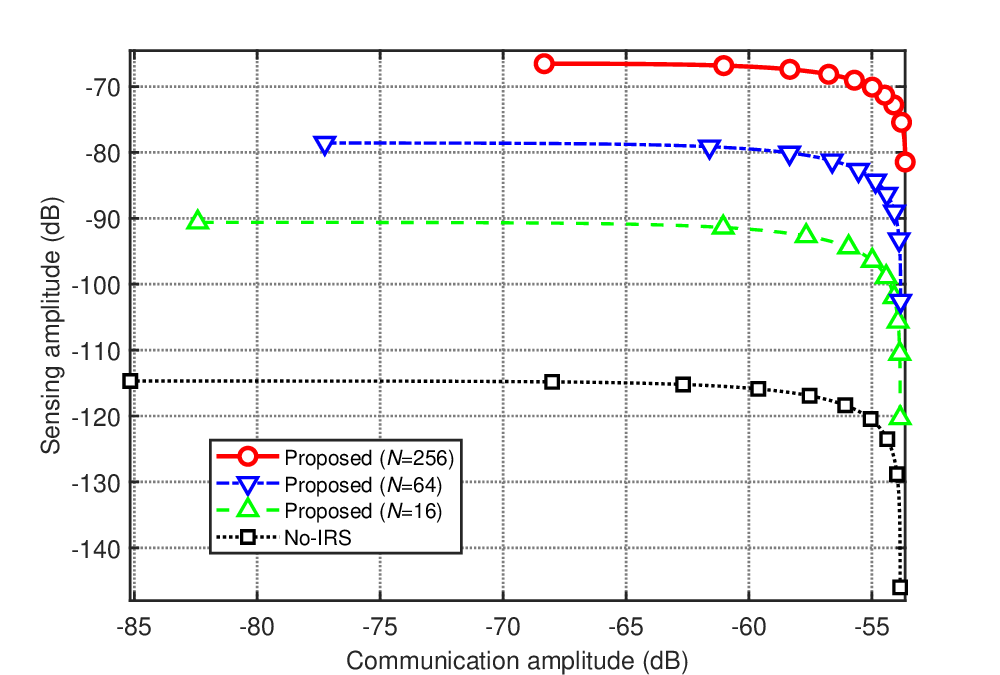} 
        \caption{$\textbf{p}_{\text{UE}} = [10,10,2]$} % 这里的文字会自动跟在 (a) 后面
        \label{fig:sub_a}
    \end{subfigure} 
    \vspace{0.5cm} % 调整两个子图之间的垂直间距 
    % --- 第二个子图 (b) ---
    \begin{subfigure}[b]{\columnwidth}
        \centering
        \includegraphics[width=0.95\columnwidth]{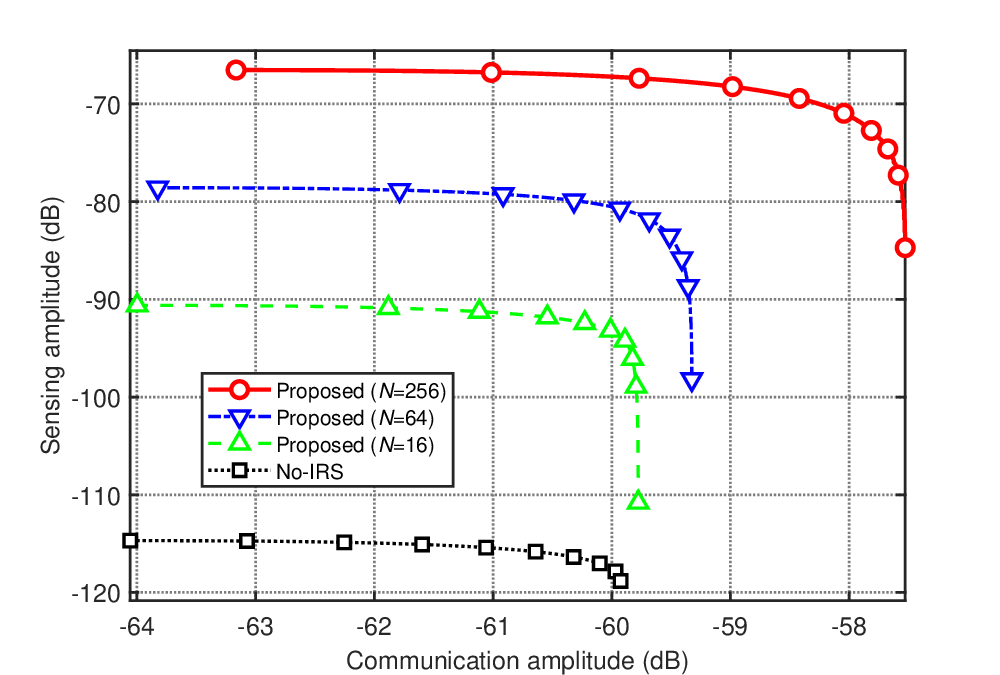}
        \caption{$\textbf{p}_{\text{UE}} = [40,40,2]$} % 这里的文字会自动跟在 (b) 后面
        \label{fig:sub_b}
    \end{subfigure} 
    % 整个大图的标题
    \caption{Pareto frontier of sensing and communication with different
the number of reflective elements.}
    \label{fig:main}
\end{figure}
\begin{figure*}[htbp] % 【重点1】加星号 *，表示跨越两栏
    \centering
\captionsetup[subfigure]{justification=centering}   
    % === 第一行 ===
    
    % 子图 (a)
    % 【重点2】宽度改成 0.48\textwidth (必须小于0.5才能并排)
    \begin{subfigure}[b]{0.45\textwidth} 
        \centering
        % 图片宽度设为 \linewidth 即可，它会自动适应上面的 0.48
        \includegraphics[width=\linewidth]{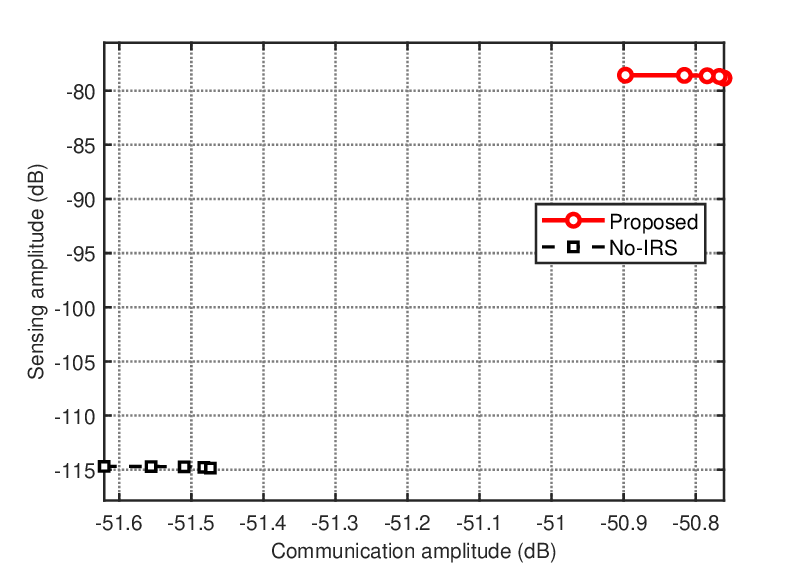} 
        \caption{$\rho_{\text{BS}}=10^\circ,\rho_{\text{IRS}}=10^\circ$}
        \label{fig:a}
    \end{subfigure}
    \hfill % 【重点3】这个命令把两个图往左右两边撑开
    % 子图 (b)
    \begin{subfigure}[b]{0.45\textwidth}
        \centering
        \includegraphics[width=\linewidth]{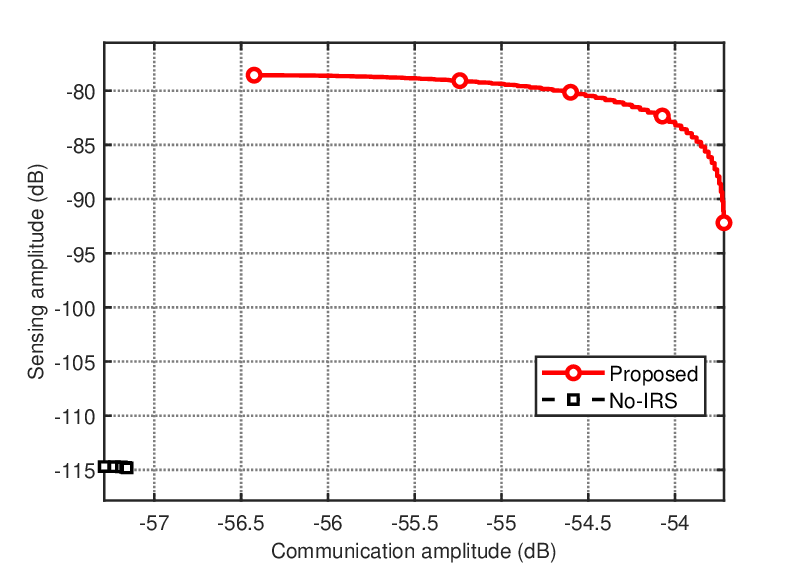}
        \caption{$\rho_{\text{BS}}=10^\circ,\rho_{\text{IRS}}=80^\circ$}
        \label{fig:b}
    \end{subfigure}
    
    \vspace{0.3cm}
    
    % === 第二行 ===
    
    % 子图 (c)
    \begin{subfigure}[b]{0.45\textwidth}
        \centering
        \includegraphics[width=\linewidth]{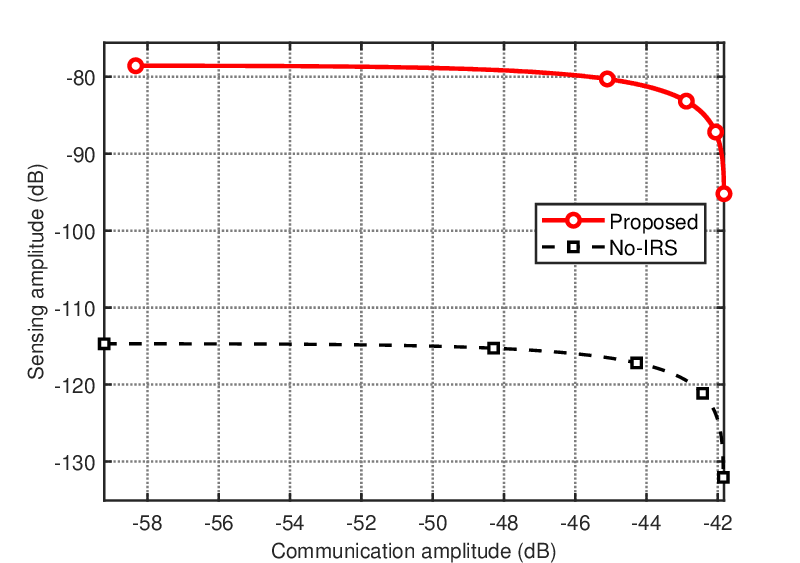} % 替换你的图片名
        \caption{$\rho_{\text{BS}}=80^\circ,\rho_{\text{IRS}}=10^\circ$}
        \label{fig:c}
    \end{subfigure}
    \hfill
    % 子图 (d)
    \begin{subfigure}[b]{0.45\textwidth}
        \centering
        \includegraphics[width=\linewidth]{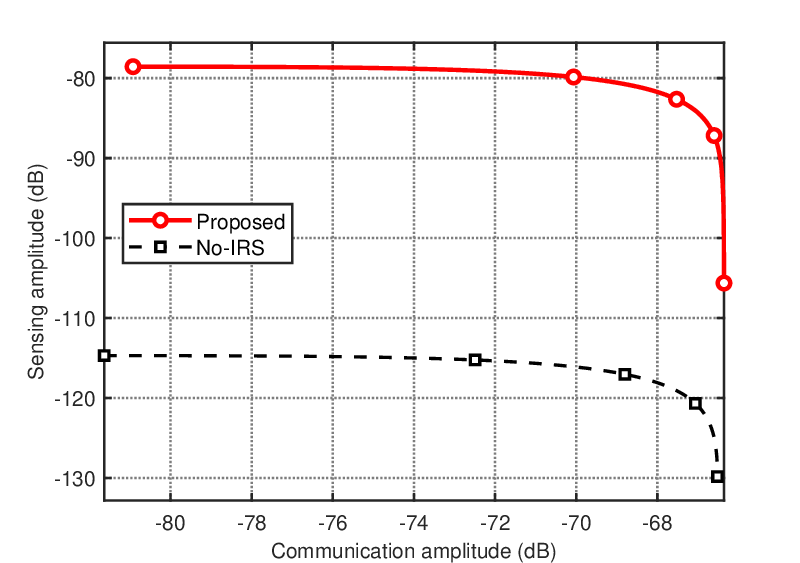} % 替换你的图片名
        \caption{$\rho_{\text{BS}}=80^\circ,\rho_{\text{IRS}}=80^\circ$}
        \label{fig:d}
    \end{subfigure}
    
    % 大图总标题
    \caption{Pareto frontier of sensing and communication with different
geometric parameters.}
    \label{fig:main_2x2}
\end{figure*}
Fig. 3 illustrates the performance trade-off under different UE locations, revealing that sensing performance is predominantly determined by the IRS scale. Specifically, enlarging the IRS to $N=256$ yields an immense sensing gain of approximately 45 dB over the ``No-IRS" benchmark. This enhancement occurs because the large IRS aperture provides crucial passive beamforming gain to compensate for the severe round-trip path loss inherent to low-RCS targets.

On the other hand, the communication gain provided by the target-mounted IRS is highly sensitive to the transceiver distance. Due to the multiplicative path loss inherent in the reflection link, the IRS is only effective when its passive beamforming gain can overcome this severe attenuation. In the short-distance scenario of Fig. 3(a) ($\mathbf{p}_{\text{UE}}=[10, 10, 2]$), the communication link is dominated by the strong direct path. Consequently, even with a large-scale IRS ($N=256$), almost no gain is observed compared to the ``No-IRS" baseline. In sharp contrast, as the user moves to the cell edge in Fig. 3(b) ($\mathbf{p}_{\text{UE}}=[40, 40, 2]$), the direct signal weakens significantly. In this regime, the array gain of the IRS becomes pivotal, where the configuration with $N=256$ yields a notable gain of approximately 3 dB over the baseline, validating the efficacy of the target-mounted IRS in extending coverage for distant users.

{\color{blue}From a physical perspective, the target-mounted IRS acts as a synthetic amplifying reflector that actively synthesizes a massive Equivalent Radar Cross-Section (ERCS) scaling with $N^2$. According to the radar equation, the sensing echo suffers from a severe $r^{-4}$ two-way path loss, which is typically a fatal bottleneck for small UAVs with weak physical RCS. However, our configuration dictates that the received echo power is proportional to $N^2 \cdot r^{-4}$. Quantitatively, increasing the reflecting elements to $N=256$ yields an overwhelming $\sim$48 dB power gain. This $N$-dependent spatial gain perfectly offsets the $r^{-4}$ attenuation, effectively extending the maximum detectable tracking range by a factor of $\sqrt{N}$ without demanding additional active transmit power. This physical mechanism solidifies the premise that the target-mounted IRS is the key enabler for tracking low-RCS targets in the low-altitude economy.}
\subsubsection{Pareto Frontier of Sensing and Communication Versus Geometric Configurations}

Fig. 4 quantitatively analyzes the Pareto frontier under distinct geometric configurations defined by spatial correlation angles $\rho_\text{BS}$ and $\rho_\text{IRS}$. When correlations at both the BS and the IRS are high (Fig. 4(a)), the achievable region shrinks to a singular optimal point (-45 dB communication, -78 dB sensing). This confirms that when the channel subspaces are highly collinear, the system effectively harvests the full $N^2$ array gain for both functions simultaneously without requiring resource splitting. In the adjacent scenario where $\rho_{\text{BS}}$ remains high but $\rho_{\text{IRS}}$ decreases, the Pareto frontier expands into a convex curve. However, the proposed scheme still provides a significant communication gain of approximately 8 dB compared to the ``No-IRS'' benchmark. This is because the joint beamforming design effectively leverages the IRS to compensate for the sensing misalignment while constructively superimposing the reflected signal with the strong direct signal, ensuring that even when maintaining a high sensing amplitude of -80 dB, the communication performance degradation is negligible compared to the conflict-free case.

The bottom row of Fig. 4 demonstrates that when BS spatial correlation is low, the active array's spatial conflict dominates the trade-off. Due to nearly orthogonal channel subspaces at the transmitter, a marginal 3 dB communication improvement (from -57 to -54 dB) causes a steep 12 dB sensing decline (from -78 to -90 dB). Nonetheless, even in the worst-case scenario (low correlations at both BS and IRS), the proposed scheme maintains a sensing amplitude above -90 dB across the Pareto frontier—outperforming the “No-IRS” benchmark (capped below -115 dB) by over 25 dB. This confirms that the target-mounted IRS is indispensable for extending the feasible region under spatially constrained direct paths.

\subsection{Tracking Accuracy and Energy Efficiency}

\subsubsection{Estimation Error Versus Time}
\begin{figure}[htbp]
    \centering
            \captionsetup[subfigure]{justification=centering}
    % --- 第一个子图 (a) ---
    % 宽度设为 1.0\textwidth 表示占满一行，这样第二个图就会自动换行到下面
    \begin{subfigure}[b]{\columnwidth}
        \centering
        % 插入图片，这里 width 控制图片实际大小
        \includegraphics[width=0.95\columnwidth]{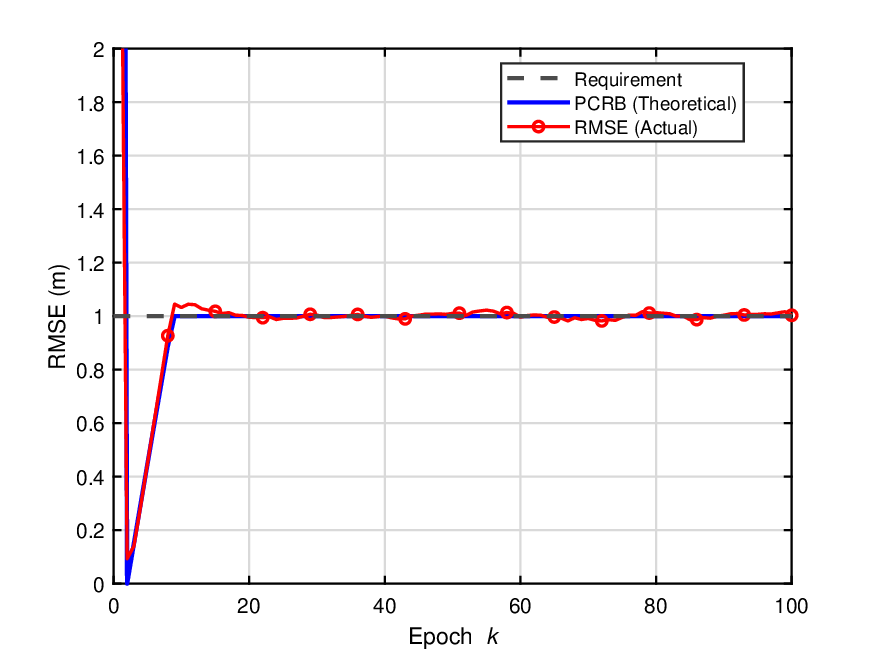} 
        \caption{Position tracking accuracy} % 这里的文字会自动跟在 (a) 后面
        \label{fig:sub_a}
    \end{subfigure} 
    \vspace{0.5cm} % 调整两个子图之间的垂直间距 
    % --- 第二个子图 (b) ---
    \begin{subfigure}[b]{\columnwidth}
        \centering
        \includegraphics[width=0.95\columnwidth]{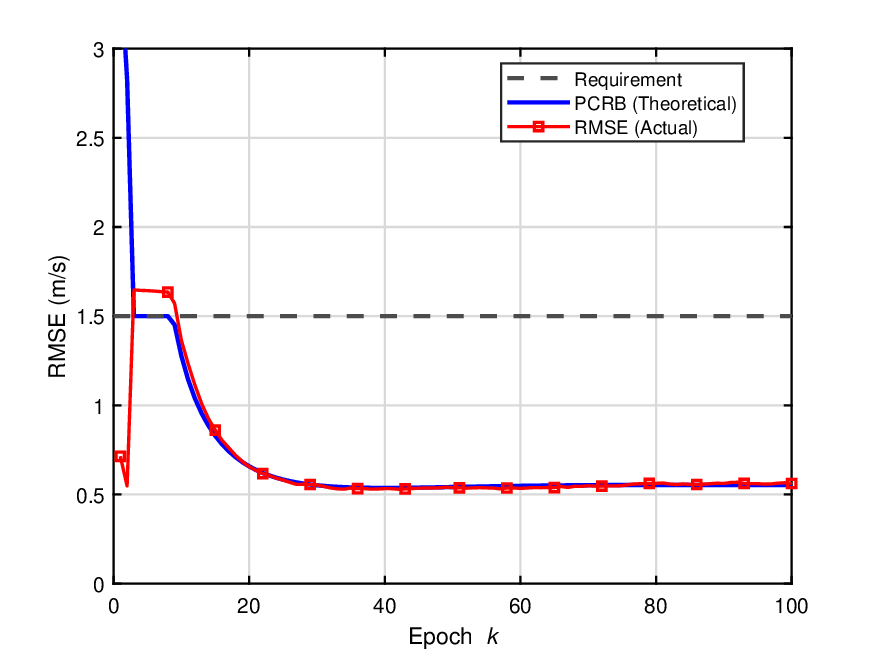}
        \caption{Velocity tracking accuracy} % 这里的文字会自动跟在 (b) 后面
        \label{fig:sub_b}
    \end{subfigure} 
    % 整个大图的标题
    \caption{Estimation error versus time for $K=100$}
    \label{fig:main}
\end{figure}

Fig. 5 illustrates the time-varying RMSE for target tracking against the PCRB benchmark. In Fig. 5(a), the location error decays rapidly, stabilizing at the centimeter level within 20 epochs and tightly converging to the theoretical bound. Critically, the steady-state error coincides exactly with the inversely derived location threshold, indicating it acts as the dominant design constraint. Consequently, the algorithm avoids unnecessary resource over-provisioning by operating precisely at the required performance limit.

In contrast, the velocity tracking performance in Fig. 5(b) reveals a distinct characteristic where the achieved error is significantly lower than its corresponding design threshold. The steady-state velocity error converges to approximately 0.05 m/s, which represents a performance margin of over 5 dB compared to the derived requirement. This phenomenon arises because the threshold derived for velocity estimation is relatively looser compared to the stringent location specification. Since the system must maintain a high SNR to satisfy the dominant location requirement, the resulting channel conditions naturally support a velocity estimation precision that far exceeds its basic baseline requirement. Consequently, the superior performance in velocity tracking is a byproduct of the system meeting the more challenging location specifications.

\subsubsection{Transmit power
Versus Time}
\begin{figure}%[htbp]
\centerline{\includegraphics[width=1\columnwidth]{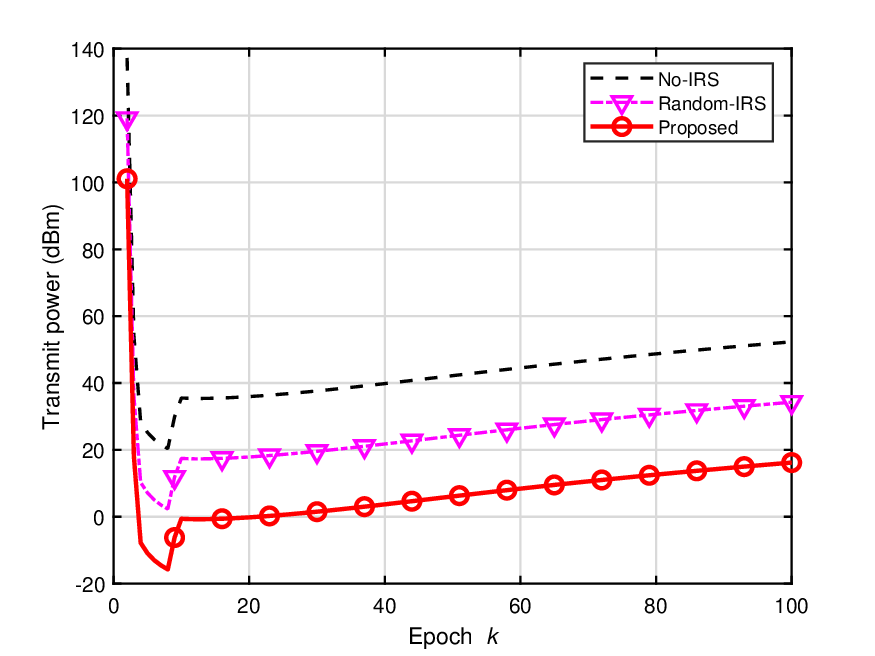}}
	\caption{Transmit power versus time for $K=100$.}\label{p1}
\end{figure}
Fig. 6 illustrates the dynamic transmit power consumption during target tracking. Initially (epochs 1-8), the power peaks at roughly 30 dBm, as the system allocates excessive transmit power to compensate for severe initial location uncertainty and potential beam misalignment. Subsequently, as the EKF converges and estimation accuracy improves, the power consumption drops and stabilizes around 22 dBm.

\section{Conclusions}
This paper has established a comprehensive framework for a target-mounted IRS-assisted ISAC system, characterized by a unified signal modeling where a dual-functional waveform is exploited for simultaneous downlink communication and 3D target tracking. We derived the PCRB to quantify the fundamental limit of the target's 3D spatial state evolution and employed the EKF for real-time state recursion. To address the inherent resource conflict, we formulated a transmit power minimization problem that adaptively selects the state update parameters and optimizes the active-passive beamforming. A low-complexity solution based on case analysis was proposed to strictly satisfy the sensing accuracy and communication QoS constraints.

Simulation results demonstrate that the proposed scheme significantly outperforms benchmark schemes. The analysis confirms that the adaptive adjustment of update parameters allows the system to operate at the optimal boundary of the Pareto frontier, significantly reducing power consumption while maintaining robust tracking performance. 

{\color{blue}Future work will focus on extending this framework to multi-target swarming scenarios by investigating sub-array partitioning and orthogonal ISAC waveforms. Furthermore, to combat severe ground clutter and multi-path interference prevalent in low-altitude environments, advanced Bayesian tracking algorithms will be integrated to ensure robust 3D tracking stability.}

\bibliographystyle{IEEEtran}
\bibliography{reference}
\end{document}